\documentclass[useAMS,usenatbib]{mn2e}
\usepackage{graphicx}

\title[Spectroscopic properties of LSBGs]
{A large sample of low surface brightness disc galaxies from the
SDSS- II. Metallicities in surface brightness bins}

\author[Liang et al.]{Y. C. Liang$^{1,2}$\thanks{E-mail:
ycliang@nao.cas.cn}, G. H. Zhong$^{1,2,3}$, F. Hammer$^{4}$,
X. Y. Chen$^{1,2,3}$,  F. S. Liu$^{5,1,2}$, 
\newauthor D. Gao$^{1,2,3}$, J. Y. Hu$^{1,2}$, L. C. Deng$^{1,2}$, B. Zhang$^{6,1,2}$  \\
 $^1$National Astronomical Observatories, Chinese Academy of Sciences,
     20A Datun Road, Chaoyang District, Beijing 100012, China\\
 $^2$Key Laboratory of Optical Astronomy, National Astronomical 
Observatories, Chinese Academy of Sciences, Beijing 100012, China \\
 $^3$Graduate School of the Chinese Academy of Sciences, Beijing 100049, China \\
 $^4$GEPI, Observatoire de Paris-Meudon, 92195 Meudon, France \\
 $^5$College of Physics Science and Technology, Shenyang Normal University, 
 Shenyang 110034, China \\
 $^6$Department of Physicals,Hebei Normal University,Shijiazhuang 050016, China \\
}
\begin{document}

\date{Accepted . Received }

\pagerange{\pageref{firstpage}--\pageref{lastpage}} \pubyear{2002}

\maketitle

\label{firstpage}

\begin{abstract}
We study the spectroscopic properties of a large sample of Low Surface 
Brightness galaxies (LSBGs)
(with B-band central surface brightness $\mu_0(B)>$22 mag arcsec$^{-2}$)
selected from the Sloan Digital Sky Survey Data Release 4 (SDSS-DR4) 
main galaxy sample. A large sample of 
 disk-dominated High Surface 
Brightness galaxies (HSBGs, with $\mu_0(B)<$22 mag arcsec$^{-2}$) 
are also selected for comparison simultaneously.
To study them in more details, these sample galaxies are
further divided into four subgroups according to $\mu_0(B)$ 
(in units of mag arcsec$^{-2}$):
vLSBGs (24.5-22.75),
iLSBGs (22.75-22.0),
iHSBGs (22.0-21.25), and
vHSBGs ($<$21.25).
The diagnostic diagram from spectral emission-line ratios shows
that the AGN fractions of all the four subgroups are small ($<$9\%).
The 21,032 star-forming galaxies 
with good quality spectroscopic
observations are further selected for studying their 
dust extinction, strong-line ratios, metallicities and 
stellar mass-metallicities relations.
The vLSBGs have lower extinction values and have less metal-rich and
 massive galaxies than the other subgroups. 
The oxygen abundances of our LSBGs are not as low as those of the 
 H~{\sc ii} regions in LSBGs studied in literature, 
 which could be because our samples are more luminous, and because of
 the different metallicity calibrations used.
 We find a correlation between 12+log(O/H) and $\mu_0(B)$ 
for vLSBGs, iLSBGs and iHSBGs
but show that 
 this could be a result of correlation between $\mu_0(B)$ and stellar 
 mass and the well-known mass-metallicity relation.
 This large sample shows that LSBGs span a wide range in metallicity
 and stellar mass, and they 
lie nearly on the stellar mass vs. metallicity and N/O vs. O/H relations of
normal galaxies. This suggests that LSBGs and HSBGs 
 have not had dramatically different star formation and chemical 
 enrichment histories. 
\end{abstract}


\section{Introduction}

Low Surface Brightness Galaxies (LSBGs)
 are important populations in the 
 field environments. However, their 
properties were seldom studied and their contributions to the galaxy population
were underestimated since they are hard to find owing to their faintness compared 
with the night sky. 
An initial quantitative study was done by Freeman (1970), who noticed that the central
surface brightness of their 28 out of 36 disc  galaxies fell within a rather 
narrow range, $\mu_0$(B)=21.65$\pm$0.3 mag arcsec$^{-2}$. This could be caused by
selection effects as pointed out by Disney (1976), and
had been previously recognized by Zwicky (1957). 

Since then, many efforts were made to search for 
LSBGs from surveys, 
and a sample of LSBGs were gathered after then. 
Specially, Impey and his colleagues
adopted the Automated Plate Measuring (APM) mechanism to scan UK
Schmidt plates and discovered 693 LSB field galaxies which forms the most
extensive catalog of LSBGs to that date (Impey et al. 1996). 
Then, their group studied the properties of this sample of LSBGs in
a series of works. 
O'Neil et al. (1997a,b in ``Texas survey") firstly found the red
LSB galaxy populations. 
Monnier Ragaigne et al. (2003a,b,c) selected a
sample of about 3800 LSBGs from the all-sky near-infrared 2MASS
survey, and then obtained the 21 cm H~{\sc i} observations and 
estimated the H~{\sc i} masses for two subsamples of 367 and 334 LSBGs. 
More related references about LSBG searching and catalogs could be found 
in Bothun, Impey, McGaugh (1997), Impey \& Bothun (1997)
and Zhong et al. (2008).

The modern Sloan digital sky survey (SDSS, York et al. 2000) 
has its great advantages to 
extend the sample of LSBGs and study their physical properties in details.
It provides to the public the homogeneous data of
photometric and spectroscopic observations for 
several hundred thousands galaxies. Therefore, it will be
very powerful and appropriate to search for a large sample of LSBGs from the
SDSS.

Kniazev et al. (2004) developed a method to search for LSBGs from
the SDSS early data release (EDR, Stoughton et al. 2002) field images and used 
the sample of Impey et. al. (1996) to test their
method. They recovered 87 same objects as in Impey et al. and 42 new LSBGs.
Bergvall et al. (2009) studied the red halos of 1510 nearly edge-on LSBGs from
SDSS. Mattsson et al. (2008) discussed the N, O abundances of a subsample of them.
Rosenbaum \& Bomans (2004) studied the large-scale environment of LSBGs
based on SDSS-EDR data, and Rosenbaum et al. (2009) extend the sample to 
Data Release 4 (DR4, Adelman-McCarthy et al. 2006) to get more detailed results.
Our group has successfully selected a large
sample of nearly face-on disc LSBGs from SDSS-DR4 main galaxy sample 
(Strauss et al. 2002).
In Zhong et al. (2008), we present their basic photometric properties
including the clear correlations of disk scalelength versus  
B-band absolute magnitude and
distance (also see the similar results of Graham \& Worley 2008 for 
 high surface brightness galaxies, HSBGs),
and the stellar populations from colors. This large
sample includes 12,282 LSBGs with $\mu_0$(B)$<$22 mag arcsec$^{-2}$, and
another 18,051 HSBGs with 
$\mu_0$(B)$>$22 mag arcsec$^{-2}$ are also selected for comparisons. 

Chemical abundance is an fundamental property of galaxy.
 The chemical properties of stars and gas within a galaxy
provide both a fossil record of its star formation history
and information on its present evolutionary status. 
Some researchers have obtained the metallicities of some H~{\sc ii} regions in
a small sample of LSBGs.
For example,
McGaugh (1994) obtained the oxygen abundances of 41
H~{\sc ii} regions in 22 LSBGs,
Roennback \& Bergvall (1995)
   obtained oxygen abundances 
   for 24 H~{\sc ii} regions in 16 blue LSBGs. 
de Blok \& van der Hulst (1998) present measurements of the oxygen abundances
of 64 H~{\sc ii} regions in 12 LSBGs,
Kuzio de Naray et al. (2004) reported the oxygen abundances of 16
H~{\sc ii} regions in 6 LSBGs.
Their results show that these H~{\sc ii} regions in the sample of LSBGs 
have low metallicities, 
the 12+log(O/H) is about 8.06 to 8.20 (in median in the samples), 
which could mean 
that the LSBGs are metal-poor and unevolved systems.

In this second paper of our series work about the large sample of 
nearly face-on LSBGs
selected from the SDSS-DR4 main galaxy sample (Zhong et al. 2008, Paper I),
 we will study their spectroscopic properties,
in particular, quantities that can
be derived from nebular emission line ratios.
We will measure nuclear activity,
 dust extinction, oxygen abundances, nitrogen to oxygen ratios,
 and examine
   the stellar mass-metallicity relations, the relations of metallicities
   vs. $\mu_0$(B) and stellar masses vs. $\mu_0$(B).
The same property parameters are also obtained for the HSBGs for comparisons.
To be in more details, the LSBGs and HSBGs are taken as a whole sample to be
divided into four subgroups according to $\mu_0$(B)
for the studies mentioned above.
 
   This paper is organized as follows. In Sect.2, 
   the sample selection criteria are given and the 
   Active Galactic Nucleus (AGN) fractions are estimated
   from BPT (Baldwin, Phillips, Terlevich 1981) diagram.
   In Sect.3, we present the dust extinction, 
   diagnostic diagram of emission-line
   ratios, oxygen abundances and log(N/O) abundance ratios of the sample
   galaxies.
  In Sect.4, we present the stellar mass-metallicity relations,
  and the relations of 12+log(O/H) vs. $\mu_0$(B), and
  stellar mass log$M_*$ vs. $\mu_0$(B) for the sample galaxies.
  The discussions are given in Sect.\ 5. 
  In Sect.\,6, we summarize and conclude the work.
 Throughout this paper, a cosmological model
with  $H_0$=70 km s$^{-1}$ Mpc$^{-1}$, $\Omega _M$=0.3 and $\Omega _\Lambda =0.7$
has been adopted.

\section[]{The sample}

The parent sample of this work is the 30,333 nearly face-on disc galaxies
selected from SDSS-DR4 main galaxy sample by Zhong et al. (2008).
Here we 
 match this sample with the emission line catalog
of SDSS-DR4 to get a subsample of star-forming galaxies
to study their spectral properties, especially metallicities.
 The fluxes of the emission-lines of the sample galaxies are taken from
 publication of the MPA/JHU collaboration (the MPA SDSS 
 website\footnote{http://www.mpa-garching.mpg.de/SDSS/},
 Kauffmann et al. 2003a,b; Brinchmann et al. 2004; Tremonti et al. 2004 etc.).
 These measurements were obtained
 from the stellar-feature subtracted spectra
 with the spectral population synthesis code of Bruzual \& Charlot (2003).
The sample selection criteria of this work are given as follows. 
To be clear, the selection criteria in Zhong et al. (2008) are also 
simply mentioned. 

\begin{figure} 
\centering
\includegraphics [width=7.5cm, height=6.5cm]{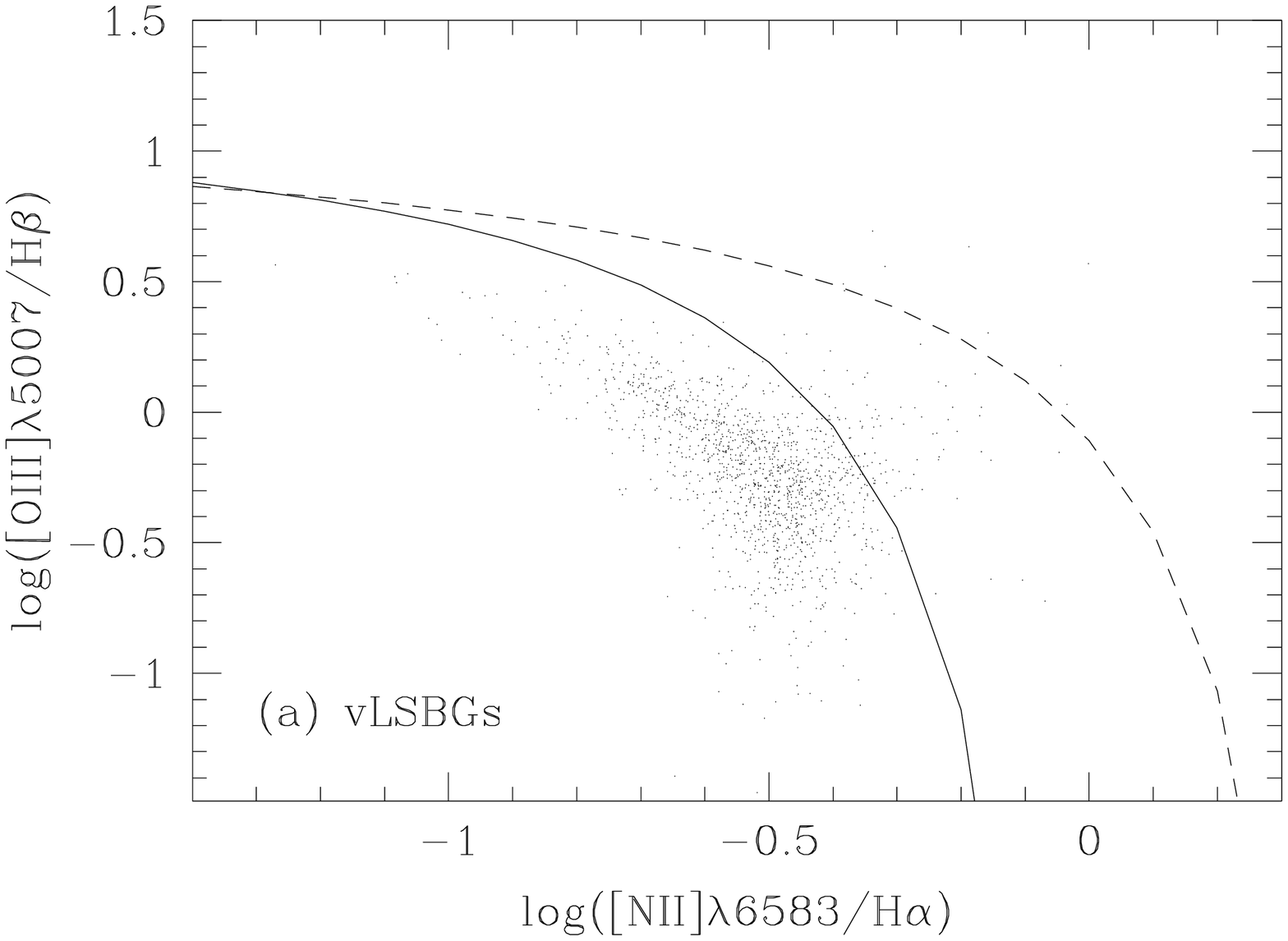} \\
\vspace{-1.8cm}
\includegraphics [width=7.5cm, height=6.5cm]{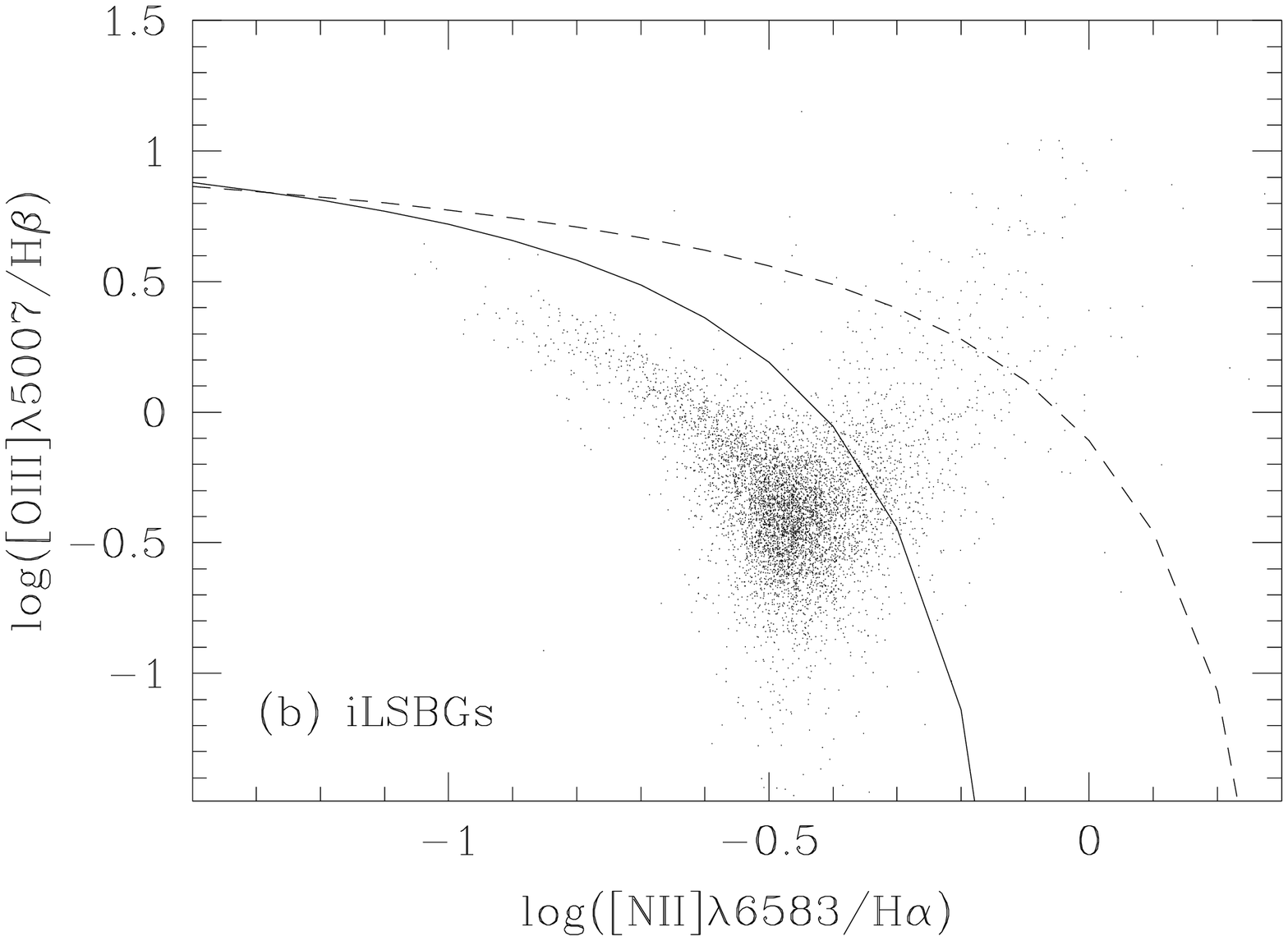} \\
\vspace{-1.8cm}
\includegraphics [width=7.5cm, height=6.5cm]{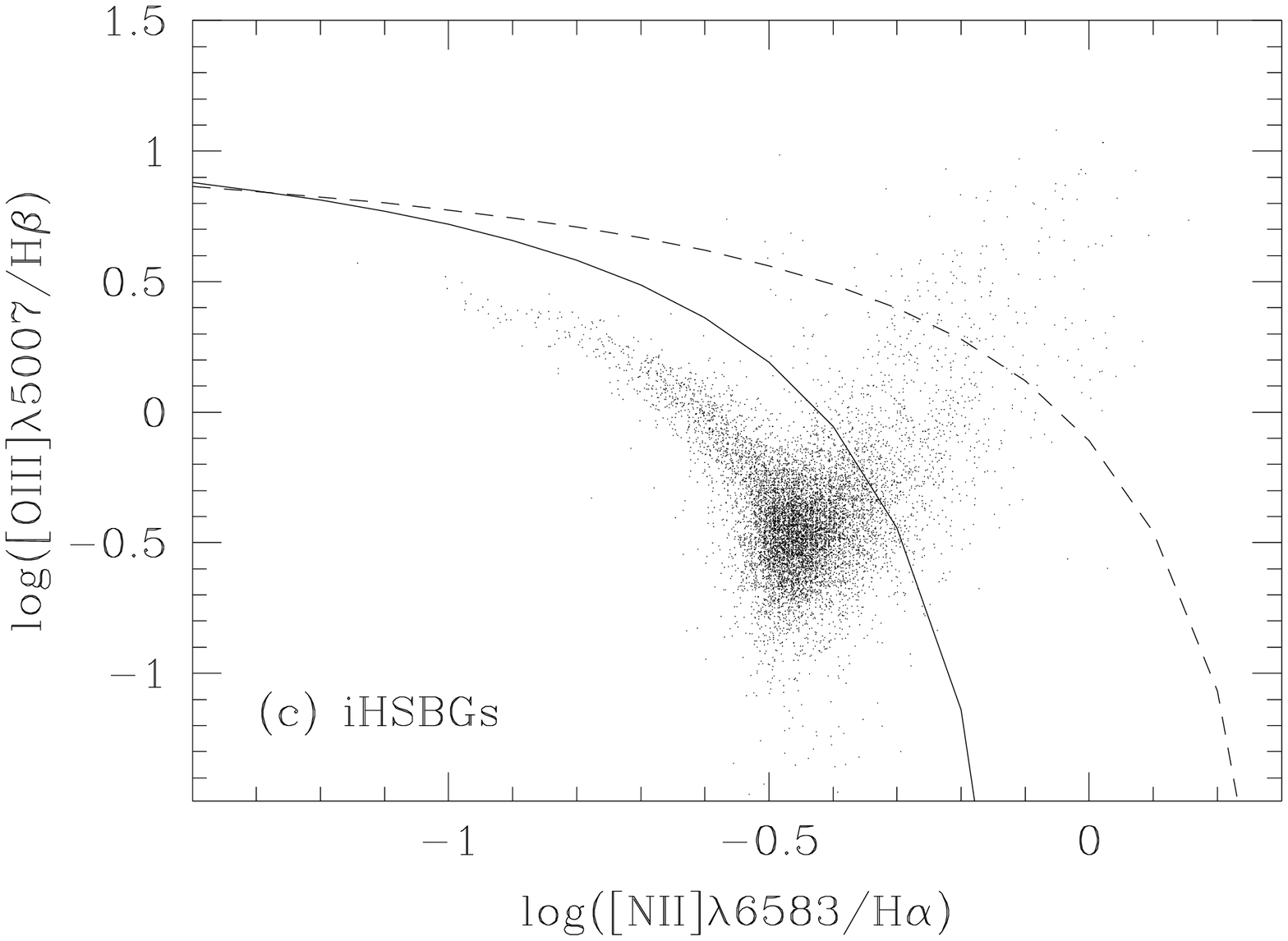} \\
\vspace{-1.8cm}
\includegraphics [width=7.5cm, height=6.5cm]{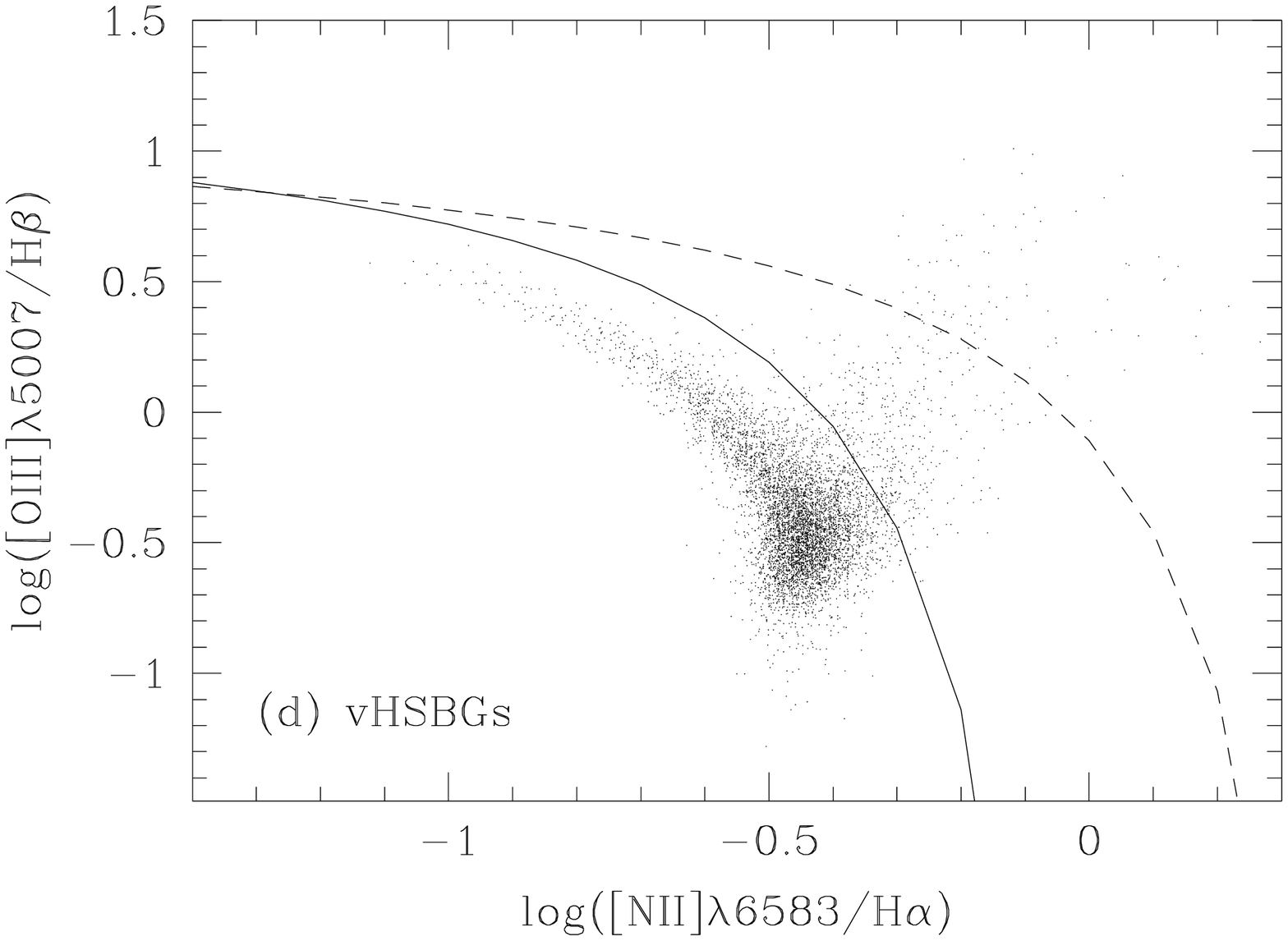} 
\vspace{-1.8cm}
\caption {The diagnostic diagrams 
for star-forming galaxies and AGNs
for the sample galaxies in 
four bins of surface brightness $\mu_0(B)$: (a) vLSBGs,
(b) iLSBGs,
(c) iHSBGs,
(d) vHSBGs.
 The diagnostic lines come from Kauffmann et al. (2003a, the solid line)
 and  Kewley et al. (2001, the dashed line). 
 The galaxies classified as AGNs fall to the upper right of the lines,
 and the star-formings fall to the lower left of the lines.
 We adopt the solid line as the diagnostic line in this work.}
\label{fig.diag}
\end{figure}

\begin{enumerate}

\item  $fracDev_r$ $<$ 0.25

 The parameter $fracDev_r$ indicates the fraction of luminosity
contributed by the de Vaucouleurs profile relative to exponential
profile in the $r$-band. 
This means the selected sample almost having an exponential 
light profile rather than a de Vaucouleurs profile. 
This can also minimize the effect
of bulge light on the disk galaxies.

\item $b/a$ $>$ 0.75

This is for selecting the nearly face-on disc galaxies, and this is  
corresponding to the inclination
$i<$41.41 degree, where $a$ and $b$ are the semi-major and
semi-minor axes of the fitted exponential disk, respectively. 

\item $M_B$ $<$ -18

This  B-band absolute magnitude cut excludes 
the few dwarf galaxies ($\sim$6\%) contained in the sample 
by the previous two steps
of selection. 
 
Up to now, 
30,333 nearly face-on disc galaxies are selected.
Taking $\mu_0$(B)=22 mag arcsec$^{-2}$ as the criterion, 
12,282 objects are classified as LSBGs and other
18,051 objects as HSBGs. More details can be found in 
Zhong et al. (2008).

\item 0.04$<z<$0.25
 
 This redshift range may ensure a covering fraction $>$20\% of 
 the galaxy light in the SDSS spectral observations, and allow us
 to get reliable metallicity estimates for the galaxies
 (Kewley et al. 2005; Liang et al. 2006a).  
 Then 28,148 (92.8\% of 30,333) of our sample galaxies 
 are within this redshift range,
 which includes 11,086 (90.3\% of 12,282) LSBGs and 
 17,062  (94.5\% of 18,051) HSBGs. 
 The aperture effects will be specially discussed in 
 Sect.\,5.

\item  Emission line selection

We require the fluxes of the emission lines have been measured 
 for 
 [O~{\sc ii}]3727, H$\beta$, [O~{\sc iii}]5007, H$\alpha$, [N~{\sc ii}]6583, 
 [S~{\sc ii}]6717,6731, and they should have higher S/N ratios
 by requiring the lines of 
 H$\beta$, H$\alpha$, and [N~{\sc ii}]6583
 detected at greater than 5$\sigma$ (following Tremonti et al. 2004
 and Liang et al. 2006a).  
 Then, it results in 22,757 (81\% of 28,148) nearly face-on disc galaxies,
 which include 7,419 (67\% of 11,086) LSBGs and 
 15,338 (90\% of 17,062) HSBGs.
 We have checked the effects of the S/N ratios of these three 
emission lines for the sample selection. If only H$\alpha$ is detected 
at greater than 5$\sigma$, the obtained LSBGs and HSBGs are 
8,005 and  15,527, respectively; if only 
[N~{\sc ii}]6583 has
S/N$>$ 5$\sigma$, the corresponding numbers are 7,916 and  15,540;
and if only H$\beta$ has
S/N$>$ 5$\sigma$, the corresponding numbers are 7,538 and 15,383.
This means that H$\beta$ plays larger role than H$\alpha$ and 
[N~{\sc ii}] in limiting the sample.
The effect of [O~{\sc iii}]5007 will be further discussed
in the later part of this section, and 
the effect of emission line selection for LSBGs will be further
discussed in Sect.\,5.

\item Four subgroups according to $\mu_0$(B)

 By following McGaugh (1996) and the discussion in Zhong et al. (2008),
we try to further divide our total sample of nearly face-on disc
galaxies (22,757) into four subgroups according to their 
B-band central surface
brightness $\mu_0$(B) (the method for calculating 
$\mu_0$(B) can be found in Sect.2.2 of Zhong et al. 2008).
We divide LSBGs into ``vLSBGs" (very Low Surface Brightness Galaxies)
 and ``iLSBGs" (intermediate Low Surface Brightness Galaxies),
 HSBGs into
 ``iHSBGs" (intermediate High Surface Brightness Galaxies) and 
 ``vHSBGs" (very High Surface Brightness Galaxies).  
 These acronyms differ slightly from those adopted by McGaugh et al. 
 (1996). 
The numbers of galaxies in the four subgroups are: 

1,364 (18\% of 7,419) vLSBGs with 22.75$<$$\mu_0(B)$$<$24.5 mag arcsec$^{-2}$,

6,055 (82\% of 7,419) iLSBGs with 22$<$$\mu_0(B)$$<$22.75 mag arcsec$^{-2}$),

9,107 (59\% of 15,338) iHSBGs with 21.25$<$$\mu_0(B)$$<$22 mag arcsec$^{-2}$, and

6,231 (41\% of 15,338) vHSBGs with $\mu_0(B)$$<$21.25 mag arcsec$^{-2}$.

Indeed, McGaugh (1996) named these four subgroups of $\mu_0$(B) bins 
as LSBGs, ISBGs, HSBGs and VHSBGs consequently.

\item Star-forming galaxies

For metallicity estimates of galaxies, only 
star-forming galaxies are selected.
 The traditional
 line diagnostic diagram [N~{\sc ii}]/H$\alpha$ vs. [O~{\sc iii}]/H$\beta$
 (BPT: Baldwin, Phillips, Terlevich 1981; Veilleux \& Osterbrock 1987) is used
 to separate the star-forming galaxies from AGNs.  We
 adopt the formula given by Kauffmann et al. (2003a, 
 the solid line in Fig.~\ref{fig.diag}):
 $\rm log([O~III]5007/H\beta)<0.61/(log([N~II]6583/H\alpha)-0.05)+1.3$, 
 to select the star-forming galaxies. 
 Kewley et al. (2001)
 also gave a diagnostic line, which often is assumed as
 the upper limit for star-forming galaxies 
 (see the dashed line in Fig.~\ref{fig.diag}).

 Finally we select 21,032 (92.4\%) star-forming galaxies 
 from the 22,757 galaxies having good quality spectral observations.
 In each of the four subgroups of surface brightness, there are
 1,299 (95.2\% of 1,324) vLSBGs,
 5,551 (91.7\% of 6,055) iLSBGs, 
 8,310 (91.2\% of 9,107) iHSBGs, and  
 5,872 (94.2\% of 6,231) vHSBGs.
 These mean that the AGN fractions of these sample galaxies are quite small,
 less than 9\%, and vLSBGs has the least fraction of AGNs, $<$5\%.
 Fig.~\ref{fig.diag} presents these results.
The AGN fraction here is much lower then that (33\%) found by Tremonti 
et al. (2004) for the whole sample of star-forming galaxies.
The reason could be that our $fracDev_r$  cut has selected against 
galaxies with bulges, and we only select the nearly face-on disc galaxies.
 The bulge
   mass is very tightly correlated with black hole mass 
   (Kormendy \& Richstone 1995; Magorrian et al. 1998).

\end{enumerate}

The median values of the B band central surface brightnesses 
$\mu_0(B)$ (in units of mag arcsec$^{-2}$) in the four sub-groups are:
22.96 for vLSBGs, 
22.28 for iLSBGs, 
21.64 for iHSBGs, and
20.86 for vHSBGs.
The corresponding mean values of them are 23.02, 22.31, 21.64 and 20.77, respectively.
 We will study these star-forming galaxies
in the following parts of this work.

 The MPA/JHU group has provided the 12+log(O/H) abundances
 for the SDSS star-forming galaxies as given in 
 Tremonti et al. (2004). 
 For our sample galaxies,
 we will adopt the oxygen abundances measured by them.
 The reason will be discussed 
 in Sect.~\ref{secn2o2r23} in details.
 Due to the number limits of the sample with reliable oxygen abundance
measurements, the numbers of our sample galaxies  
in each of the four subgroups are then reduced to  
601 (46\% of 1,299) of vLSBGs, 
2,630 (47\% of 5,551) of iLSBGs, 
5,517 (66\% of 8,310) of iHSBGs, and
5,317 (90\% of 5,872) of vHSBGs. 

 The reason for the large fraction of reduced numbers here could be 
mainly due to the S/N cut of [O~{\sc iii}]5007$>$ 5$\sigma$, which
 probably was used in producing the MPA/JHU metallicity catalog,
 although was not used by Tremonti et al. (2004).
 We have tried to check this guess.  
 In criterion (v), when we only consider [O~{\sc iii}]5007$>$ 5$\sigma$
 as the limiting for selecting the sample from S/N ratio,
 we will obtain 3,440 LSBGs and  11,463 HSBGs, which are much 
 smaller than only considering other three emission lines.
 Furthermore, if we consider all the four strong emission lines 
 H$\beta$, H$\alpha$, [N~{\sc ii}]6583 and [O~{\sc iii}]5007
 detected at greater than 5$\sigma$, 
 the resulted sample will be 3,332 LSBGs and 11,386 HSBGs, 
 which are similar to what obtained by only considering 
 [O~{\sc iii}]5007$>$ 5$\sigma$ as given above. This means, indeed, 
 [O~{\sc iii}]5007 plays largest role in limiting the sample.
 These samples (3,332 and 11,386) are further classified as
 3,014 (90.4\%) star-forming LSBGs and
 10,439 (91.6\%) star-forming HSBGs, which still
 suggest the small AGN fraction ($<$10\%) of the sample galaxies.
 Then, we divided these star-forming galaxies into four subgroups 
 following their surface brightness. There are
 567 vLSBGs, 2,447 iLSBGs, 5,181 iHSBGs and 5,258 vHSBGs.
 These four values are quite similar to the numbers of 601, 2,630, 5,517 and 5,317
 given in last paragraph, 
 which are obtained by matching the samples from criteria (i)-(vii) 
 with the metallicity catalog
 of MPA/JHU, although very slight differences still exist.
 
 The stellar masses of the sample galaxies are taken from the MPA/JHU
 database as well (Kauffmann et al. 2003b, Gallazzi et al. 2005). 
 To estimate stellar masses of the galaxies,
 they firstly estimated the dust-correction to the observed
 $z$-band magnitude of the galaxy, and then
 the stellar mass was computed by multiplying the 
 dust-corrected luminosity of the galaxy by the stellar mass-to-ligh ratio
 predicted by model.
 In their work, the M/L ratios are estimated by also considering the spectral
   features. But nearly all other methods use colors to estimate M/L.
 Due to the limits of reliable stellar mass estimates, our sample
 galaxies having both metallicities and stellar masses are then reduced to 
597 of vLSBGs,
2,609 of iLSBGs, 
5,486 of iHSBGs, and 
5,291 of vHSBGs.

All the corresponding numbers of sample galaxies and their median values of
the property parameters are presented in Table~\ref{tab1}.

\begin{table}
\begin{center}
\caption{Basic information of the sample galaxies in the four subgroups 
with $\mu_0(B)$. All the estimates of parameters are in median.
 }
\label{tab1}
\scriptsize
\begin{tabular}{lcccc}
\hline
              &  vLSBG      &    iLSBG     &   iHSBG      &    vHSBG     \\  \hline
$\mu_0(B)$    & 24.5-22.75  & 22.75-22.0   & 22.0-21.25  &  $<$21.25   \\
All galaxies  &  1,364      & 6,055       & 9,107       &   6,231    \\
 Star forming &  1,299      & 5,551       & 8,310       &  5,872     \\
 AGN fraction &  4.8\%      &  8.3\%       & 8.8\%       & 5.8\%      \\
 A$_V$  & 0.46      &  0.63  &     0.76  & 0.83   \\    \hline
with O/H    &   &    &   &     \\
 Star forming &  601      & 2,630       & 5,517       &  5,317     \\ 
 12+log(O/H)  &   8.77   &  8.94    &   9.03 &  9.06 \\
 log(N/O)      &  -1.05   & -0.94    &   -0.85  & -0.83   \\  \hline
with stellar mass    &   &    &   &     \\
 Star forming &  597      & 2,609       & 5,486       &  5,291     \\ 
 log(M$_*$)   &    9.55   &  9.91    &    10.21   & 10.29    \\
 light\_fraction   &   0.12   &  0.15    &    0.21   & 0.30    \\  \hline
\end{tabular} 
\end{center}
$Note: $ {\footnotesize Raw~(1)-(5) consequently refer to the ranges of $\mu_0(B)$
in units of mag arcsec$^{-2}$, 
the numbers of all the sample galaxies;
the numbers of star-forming galaxies;
AGN fractions 
and median values of dust extinction.
Raw~(7)-(9) refer to the numbers of the galaxies having
oxygen abundance estimates, 
the median values of oxygen abundances and log(N/O) consequently. 
Raw~(11)-(13) refer to the numbers of the galaxies having both
oxygen abundance and stellar mass estimates, 
median values of stellar masses and light fractions in the fiber observations consequently.}
\end{table}

\section{Dust extinction and nebular abundances}

We study the dust extinction, strong emission-line ratios, 
12+log(O/H) abundances and log(N/O) abundance ratios 
of our star-forming sample galaxies in four bins of 
$\mu_0(B)$ in this section.

\subsection{Dust extinction}

  The dust extinction inside the star-forming galaxies are derived using the
Balmer line ratio H$\alpha$/H$\beta$: assuming case B recombination, with a
density of 100\,cm$^{-3}$ and a temperature of 10$^4$\,K, and then the predicted intrinsic
ratio of H$\alpha$/H$\beta$ is 2.86 (Osterbrock 1989),
with the relation of 
$
(\frac{I_{H\alpha }}{I_{H\beta }})_{obs}  = 
(\frac{I_{H\alpha_0}}{I_{H\beta_0}})_{intr}10^{-c(f(H\alpha )-f(H\beta ))}.
$
Using the average
interstellar extinction law given by Osterbrock (1989), we have 
$f({\rm H}\alpha)-f({\rm H}\beta)$=$-0.37$. Then, the extinction parameter $A_V$ 
are calculated following Seaton (1979): $A_V=E(B-V)R=\frac{cR}{1.47}$ (mag).
$R$= 3.1 is the ratio of the total to the selective extinction at $V$. The
emission-line fluxes of the sample galaxies have been corrected for this
extinction. 

\begin{figure} 
\centering
\includegraphics [width=6.5cm, height=5.5cm]{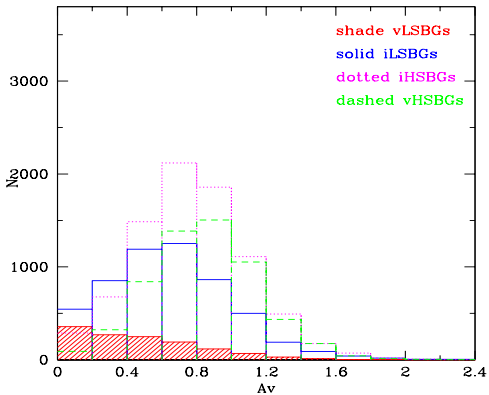} 
\caption {Histogram distributions of dust extinction Av of 
the star-forming sample galaxies in four bins of $\mu_0(B)$,
the median values are: 
0.46 for vLSBGs,  
0.63 for iLSBGs, 
0.76 for iHSBGs, 
0.83 for vHSBGs.}
\label{fig.Av}
\end{figure}

 The histogram of A$_V$ of the star-forming sample galaxies 
 with bins of 0.2 is given in Fig.~\ref{fig.Av}.
 The median values of A$_V$ (in units of mag) 
 in the four subgroups are
  0.46, 0.63,  0.76 and 0.83 for vLSBGs, iLSBGs,
  iHSBGs and vHSBGs, respectively (see Table~\ref{tab1}).
  The mean values of A$_V$ are 0.47, 0.65,  0.77 and 0.84, respectively.
  It shows that the vLSBGs have lower dust extinction than other three
  subgroups, and it is about 0.2, 0.3, 0.4 mag lower than those of the iLSBGs,
  iHSBGs and vHSBGs, respectively. 
   If [O~{\sc iii}]5007$>$ 5$\sigma$ is
  further considered in criterion (v) in Sect.\,2, 
  the selected galaxies will have 
  median values of A$_V$ as 0.36, 0.54, 0.70 and 0.82 for vLSBGs, iLSBGs,
  iHSBGs and vHSBGs, respectively. Similar to the values above. 

\subsection{Oxygen abundances}
\label{secn2o2r23}

R$_{23}$ (=([O~{\sc ii}]3727+ [O~{\sc
 iii}]4959,5007)/H$\beta$) is an indicator of metallicity of galaxy.
However, one defect of $R_{23}$ is that it results in double-valued
 12+log(O/H) abundances, which are for the metal-rich (upper) and metal-poor
 (lower) branches, respectively, with the transition  around 
 12+log(O/H)$\sim$8.4 (log$R_{23}$$\sim$0.8).
 The reason is that, at metal-poor environments, the forbidden lines scale
 roughly with the chemical abundances, while at metal-rich environments, the
 nebular cooling is dominated by the infrared fine-structure lines and the
 electron temperature becomes too low to collisionally excite the optical
 forbidden lines.
 
 Some other strong-line ratios can break this degeneracy and could also
 trace the metallicities of the galaxies,
 such as 
 [N~{\sc ii}]$\lambda$6583/H$\alpha$,
 [O~{\sc iii}]$\lambda$5007/[N~{\sc ii}]$\lambda$6583,
 [N~{\sc ii}]$\lambda$6583/[O~{\sc ii}]$\lambda$3727,
 [N~{\sc ii}]$\lambda$6583/[S~{\sc ii}]$\lambda$$\lambda$6717,6731,
 [S~{\sc ii}]$\lambda$$\lambda$6717,6731/H$\alpha$,
 and [O~{\sc iii}]$\lambda$$\lambda$4959,5007/H$\beta$.
 Liang et al. (2006a, also Nagao et al. 2006) and 
 Kewley \& Dopita (2002) study these line ratios 
 as metallicity indicators 
 from a large sample of SDSS galaxies and photoionization models,
 respectively.
 Among these line ratios, [N~{\sc ii}]$\lambda$6583/[O~{\sc ii}]$\lambda$3727 
 was confirmed as the best metallicity calibration because it shows a monotonical
 increasing following
 the increasing metallicity and less scatter than other line ratios
 due to its independence on ionization parameter.
 However, dust extinction must be estimated properly before
 using this indicator because the blue line [O~{\sc ii}] 
 is affected much by dust extinction and is far
 from [N~{\sc ii}] in wavelength.
 Kewley \& Ellison (2008) also use  [N~{\sc ii}]/ [O~{\sc ii}] to break the R$_{23}$
 degeneracy for their SDSS sample. Their Fig.8 shows that the division between the 
 R$_{23}$ upper and lower branches occurs at log([N~{\sc ii}]/ [O~{\sc ii}])$\sim -$1.2.
 But McGaugh (1994) adopted log([N~{\sc ii}]/ [O~{\sc ii}])$\sim -$1.0 as  the transition
 limit.
 
\begin{figure} 
\centering
\includegraphics [width=7.5cm, height=6.5cm]{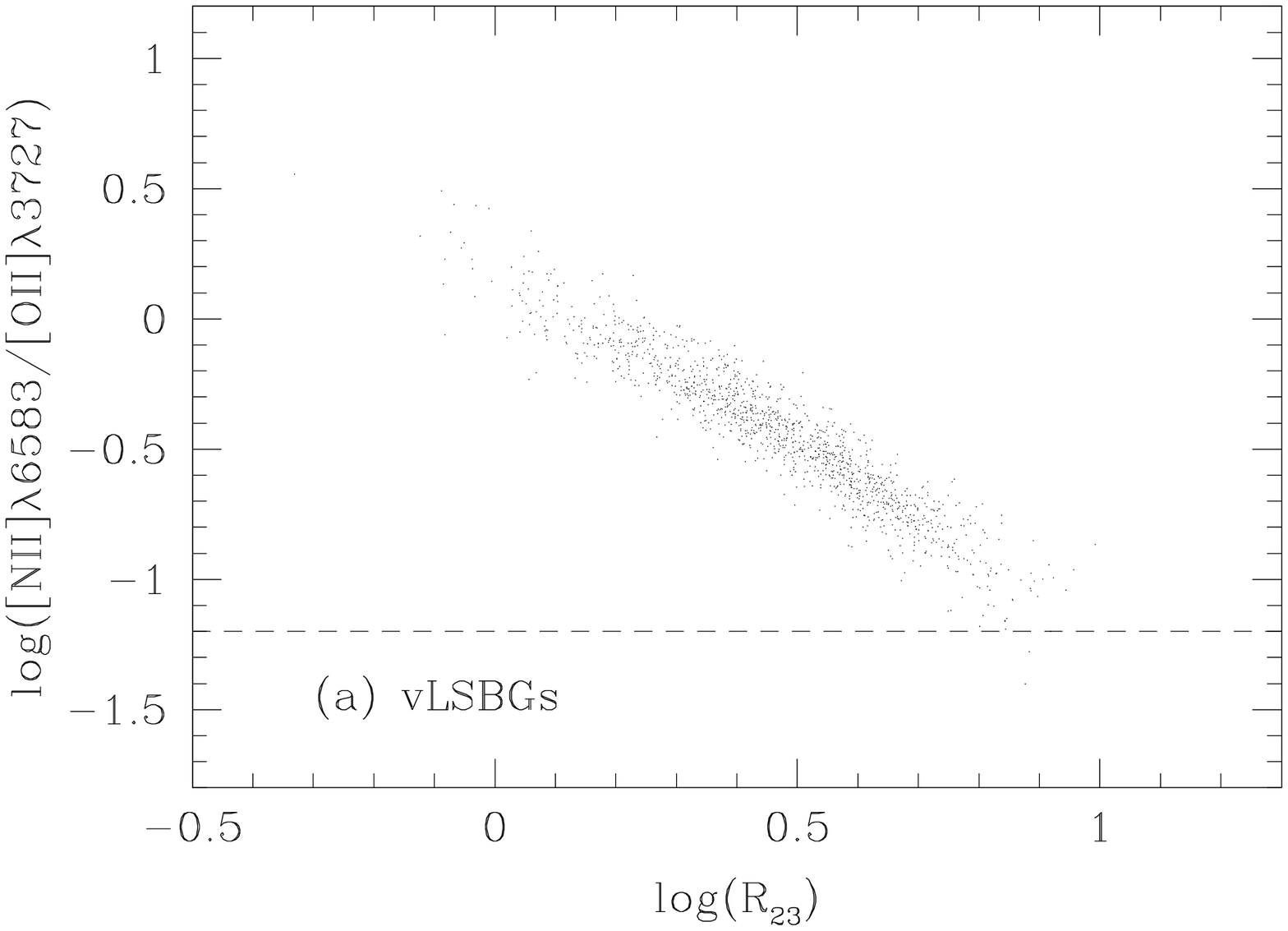} \\
\vspace{-1.8cm}
\includegraphics [width=7.5cm, height=6.5cm]{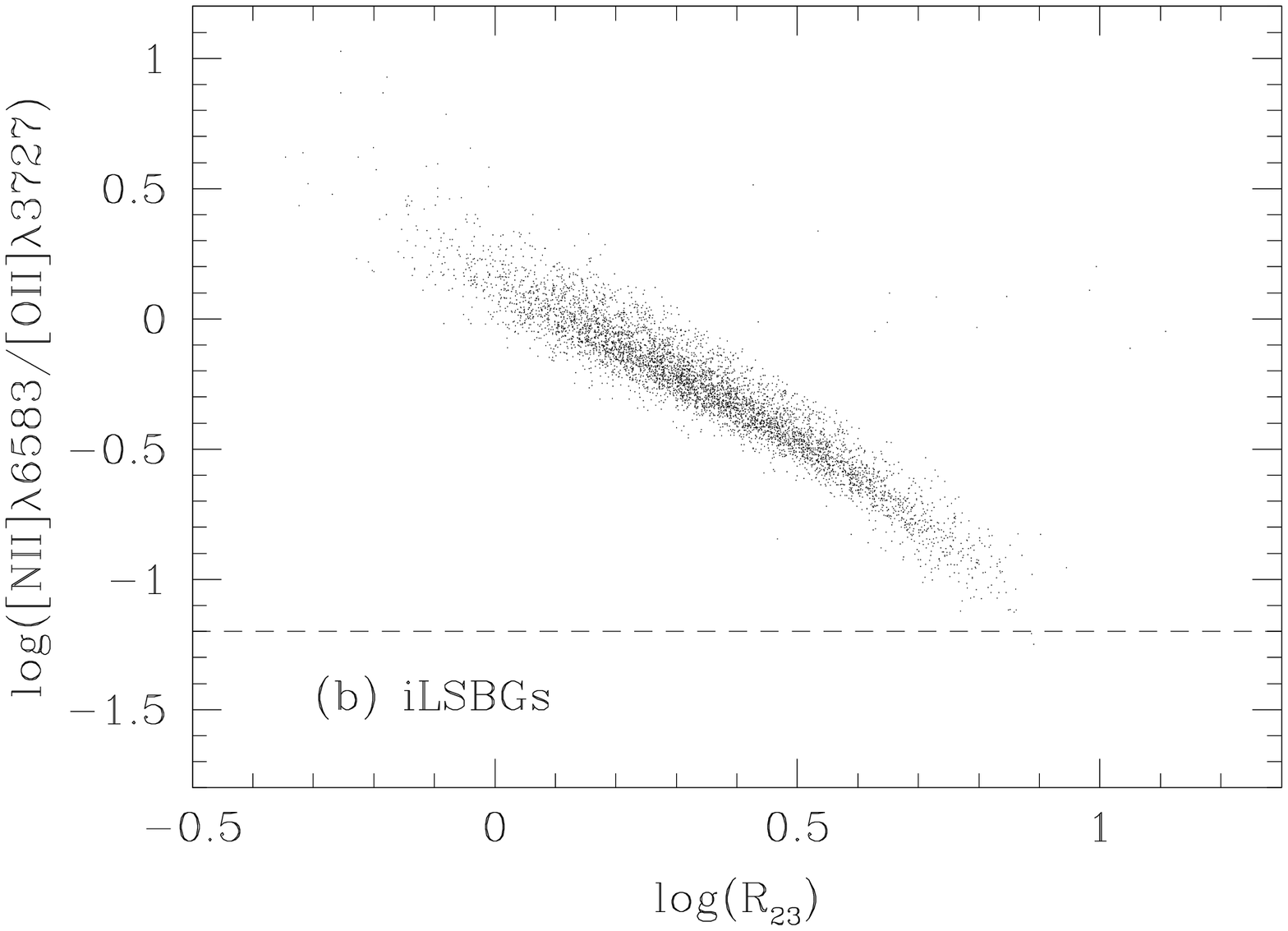} \\
\vspace{-1.8cm}
\includegraphics [width=7.5cm, height=6.5cm]{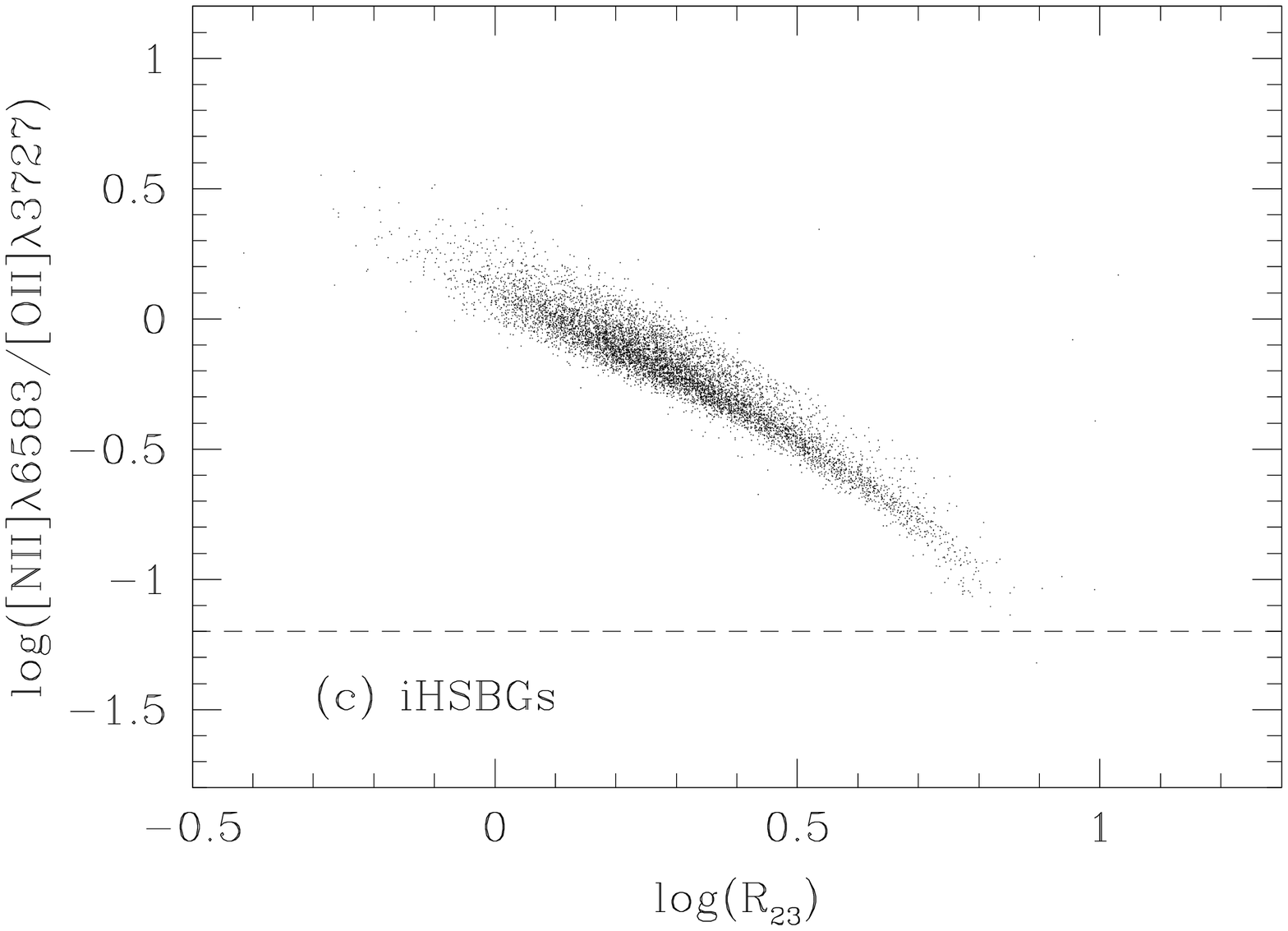} \\
\vspace{-1.8cm}
\includegraphics [width=7.5cm, height=6.5cm]{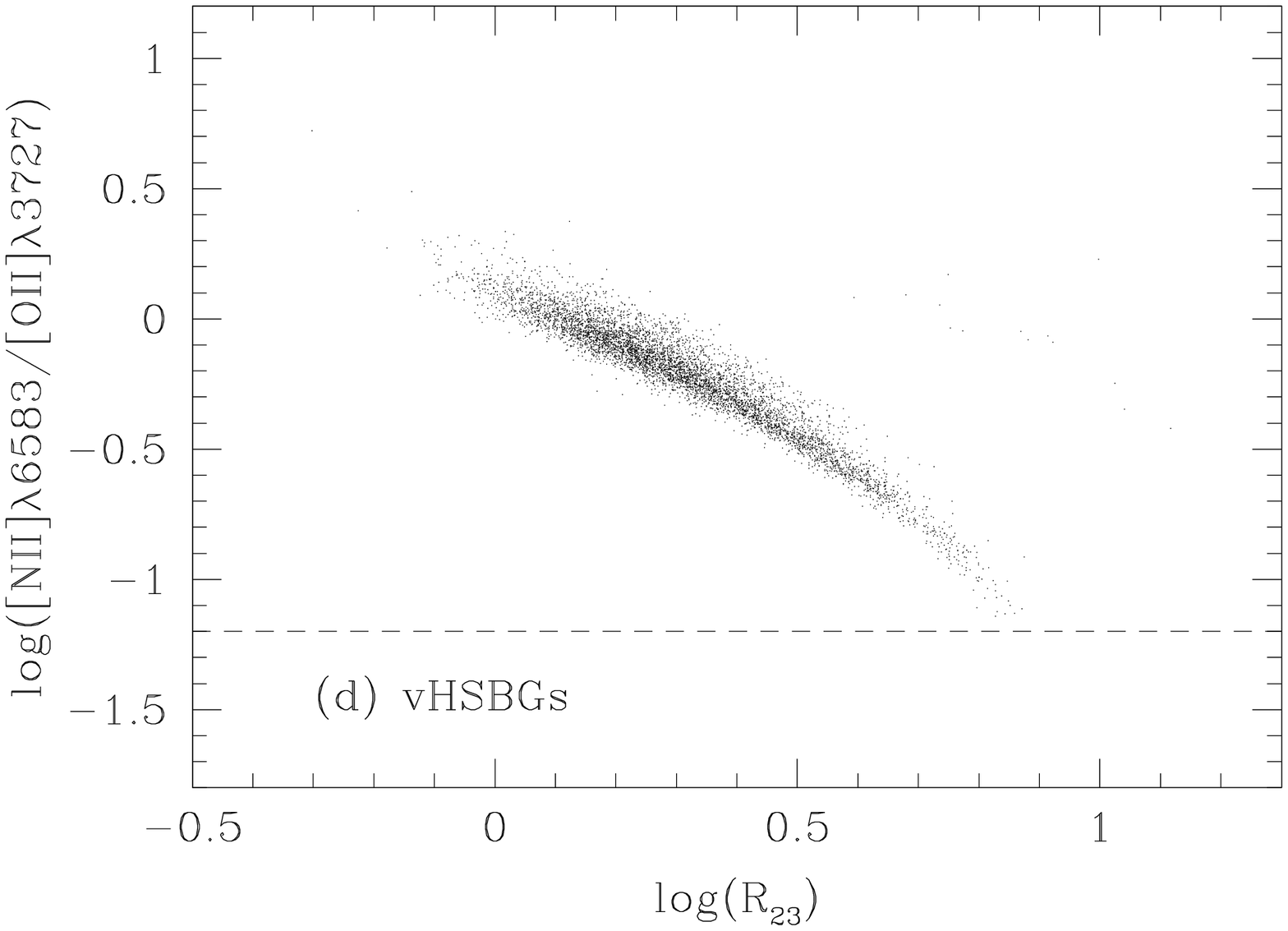} 
\vspace{-1.8cm}
\caption {The relationships of log([N~{\sc ii}]/[O~{\sc ii}]) vs. log($R_{23}$)
for the star-forming sample galaxies in four bins of $\mu_0(B)$:
(a) vLSBGs; (b) iLSBGs; (c) iHSBGs; (d) vHSBGs. Most of them have 
log([N~{\sc ii}]/ [O~{\sc ii}])$>-$1.2. }
\label{fig.N2O2R23}
\end{figure}

 We present our sample galaxies in the plot of  
 log([N~{\sc ii}]/ [O~{\sc ii}]) vs. log(R$_{23}$) in Fig.~\ref{fig.N2O2R23}.
 It shows that most of our sample galaxies have log([N~{\sc ii}]/ [O~{\sc ii}])$>-$1.2,
 meaning that they should belong to the upper branch of metallicity.
 This could hint that  
 our sample galaxies are not that metal-poor as  
 the H~{\sc ii} regions in the small samples of LSBGs studied by 
 McGaugh (1994) (their Fig.3),
 Roennback \& Bergvall (1995),
 de Blok \& van der Hulst (1998) and Kuzio de Naray et al. (2004) etc.
 They found the 12+log(O/H) of those H~{\sc ii} regions in LSBGs
 are 8.06 to 8.20 (the median values of their samples).
 More discussions will be given in Sect.5.2.

The strong-line ratios $R_{23}$ can be used to estimate the oxygen
abundances of the galaxies. 
The formula of Tremonti et al. (2004, their Eq.(1)) is used here,
which is appropriate for upper branch of metallicity, since
Fig.~\ref{fig.N2O2R23} has shown that 
most of our sample galaxies belong to the upper-branch.
However, in this study, for the oxygen abundances of our sample galaxies
 we prefer to use the values provided by the MPA/JHU group as Tremonti et al. (2004),
 rather than the $R_{23}$-based ones. 
 We have checked and found that for the galaxies having relative low
metallicities, such as those with 12+log(O/H)(T04)$<$8.6, 
the $R_{23}$ method will overestimate their abundances by about 0.1-0.3\,dex. This is 
more obvious for vLSBGs and iLSBGs, and less obvious for iHSBGs and vHSBGs.
 The reason could be that $R_{23}$ is not a good metallicity
  indicator for galaxies in the ``turn around region" (log($R_{23}$) $>$ 0.7;
  12+log(O/H) = 8.2 - 8.6).  First there is the issue of having to
  decide which branch to use. Second the relation between $R_{23}$ and
  metallicity becomes very steep in this region, hence small errors in $R_{23}$ 
  will result in large errors in O/H.  The Tremonti's oxygen abundances are free of these
  problems and more robust to observational errors since they use many
  lines.
 
\begin{figure} 
\centering
\includegraphics [width=6.5cm,height=5.5cm]{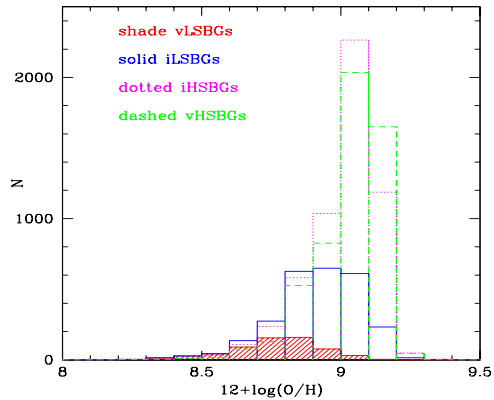} 
\caption {Histogram distributions of 12+log(O/H) of the star-forming 
sample galaxies in four bins of $\mu_0(B)$: 
the median values are
8.77 for vLSBGs,
8.94 for iLSBGs, 
9.03 for iHSBGs, 
9.06 for vHSBGs. }
\label{fig.his.oh}
\end{figure}

 Figure~\ref{fig.his.oh} shows the 
histograms distributions of the abundance values of the galaxies 
in the four subgroups, where
the median 12+log(O/H) values are
8.77 for vLSBGs,  
8.94 for iLSBGs, 
9.03 for iHSBGs, 
9.06 for vHSBGs (see Table~\ref{tab1}).
The mean values are 8.77, 8.92, 9.01 and 9.03, respectively.
This shows that the vLSBGs have the lowest metallicities among the four subgroups,
their 12+log(O/H) abundances are 0.3\,dex lower 
than those of vHSBGs and vHSBGs, 
and  0.17\,dex lower than that of iLSBGs.

However,
these results directly show that our vLSBGs and iLSBGs are not as metal-poor
as the H~{\sc ii} regions in a small sample of LSBGs studied by McGaugh (1994) 
and some following works,
which are 
8.06 to 8.20 in median of their 12+log(O/H). 
The main reasons for such difference could be the different types of galaxies 
we studied (most of theirs are dwarfs and ours are much luminous),
the different calibrations to derive the oxygen abundances
(they used the lower branch $R_{23}$ formula 
and our oxygen abundances are provided by Tremonti et al.
by using the Bayesian approach), 
and the different photoionization models used
to estimate oxygen abundances (they used McGaugh 1991 which results 
in lower oxygen abundances than what used by Tremonti et al.). 
We will discuss these carefully in Sect.5.2 in details.

 \subsection{The nitrogen to oxygen ratios}
 
 It is possible to estimate the N abundances of galaxies from strong 
optical emission lines, such as [N~{\sc ii}]6583, [O~{\sc ii}]3727 and 
log$R_{23}$,
which can help to understand the origin of nitrogen.

The basic nuclear mechanism to produce nitrogen is  
from CNO processing of oxygen and carbon in hydrogen burning.
If the ``seed" oxygen
and carbon are those incorporate into a star at its formation and a
constant mass fraction is processed, then the amount of nitrogen produced
is proportional to the initial heavy-element abundance, and the nitrogen
synthesis is said to be ``secondary". If the oxygen and carbon are produced
in the star prior to the CNO cycling (e.g. by helium burning in a core,
followed by CNO cycling of this material mixed into a hydrogen-burning
shell), then the amount of nitrogen produced may be fairly independent of
the initial heavy-element abundance of the star, and the synthesis is said
to be  ``primary" (Vila-Costas \& Edmunds 1993). In general, then, primary
nitrogen production is independent of metallicity, while secondary
production is a linear function of it. 
The N/O ratio as a function of O/H is       
the basic method to study the N abundance of galaxies.
It could be sensitive to the ratio of intermediate-mass to massive stars
since nitrogen mainly come from the intermediate-mass stars and oxygen
mostly come from the massive ones.

We use the formula given by Thurston et al. (1996) to
estimate the electron temperature in the [N~{\sc ii}] emission region
($t_{[\rm NII]}$ = $t_{II}$, in units of 10$^4$K) from log$R_{23}$:
\begin{equation}
  t_{II}=6065+1600x+1878x^2+2803x^3,
\end{equation}
where $x$=log(R$_{23}$).

Then, log(N/O) values are
estimated from the ([N~{\sc ii}]
$\lambda\lambda{6548},{6583}$)/([O~{\sc ii}] $\lambda {3727}$)
emission-line ratio and $t_{II}$ (=$t_{[NII]}$) 
by assuming $\frac{\rm N}{\rm O}=\frac{\rm N^{+}}{\rm
O^{+}}$, and using the convenient formula based
upon a five-level atom calculation given by Pagel et al. (1992) 
and Thurston et al. (1996):
\begin{eqnarray}
&\rm log ({{N^+}\over {O^+}})& = {\rm log [{{[NII]6548,6583}\over
  {[OII]3727,3729}}]+0.307   -0.02log} t_{II}  \nonumber \\
 &  & -{{0.726}\over {t_{II}}}.
\end{eqnarray}
 We consider that the flux of [N~{\sc ii}]$\lambda{6548}$ is equal to one-third of that of 
the [N~{\sc ii}]$\lambda{6583}$ in the calculations. 

\begin{figure} 
\centering
\includegraphics [width=7.5cm, height=6.5cm]{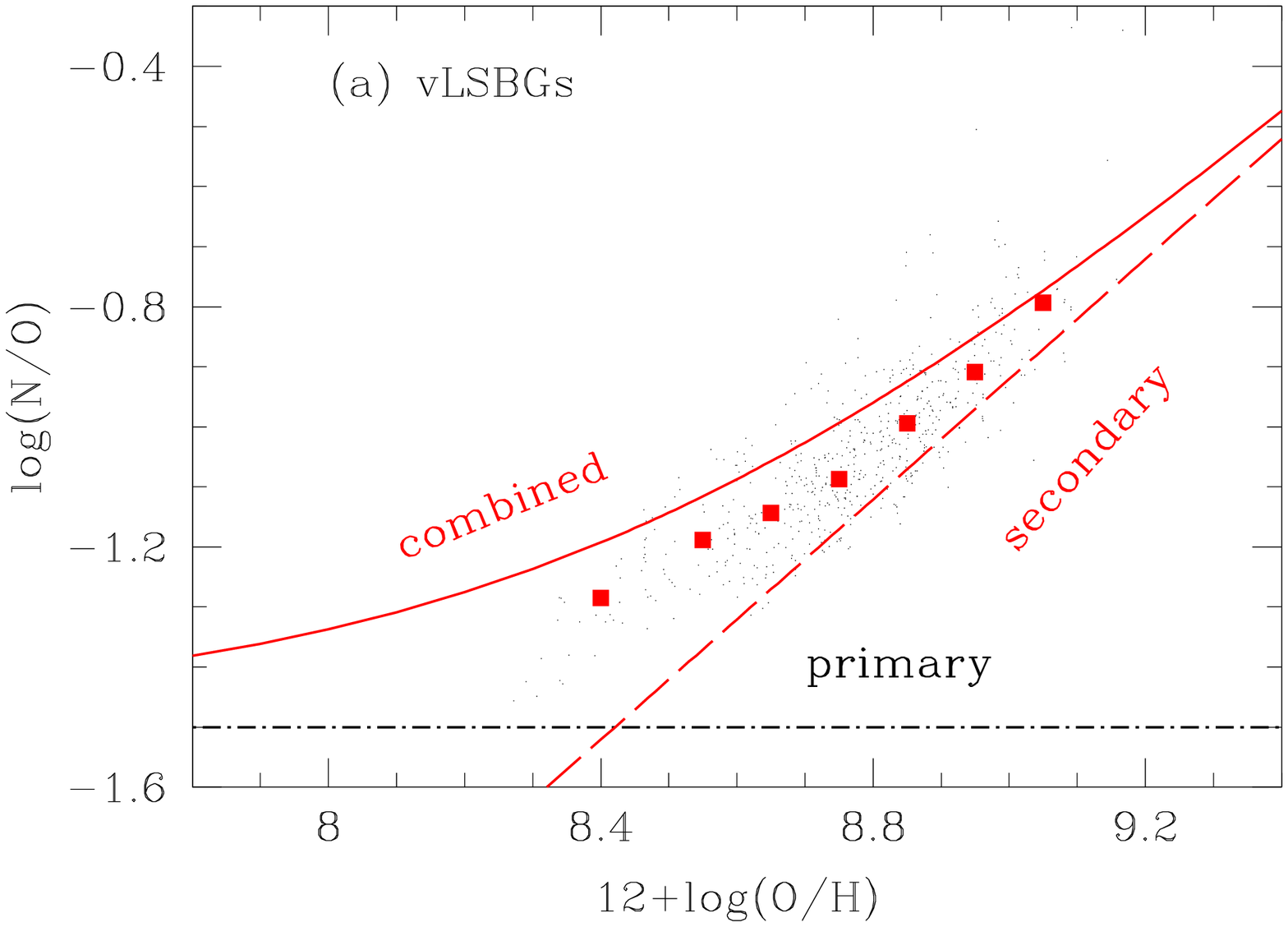} \\
\vspace{-1.8cm}
\includegraphics [width=7.5cm, height=6.5cm]{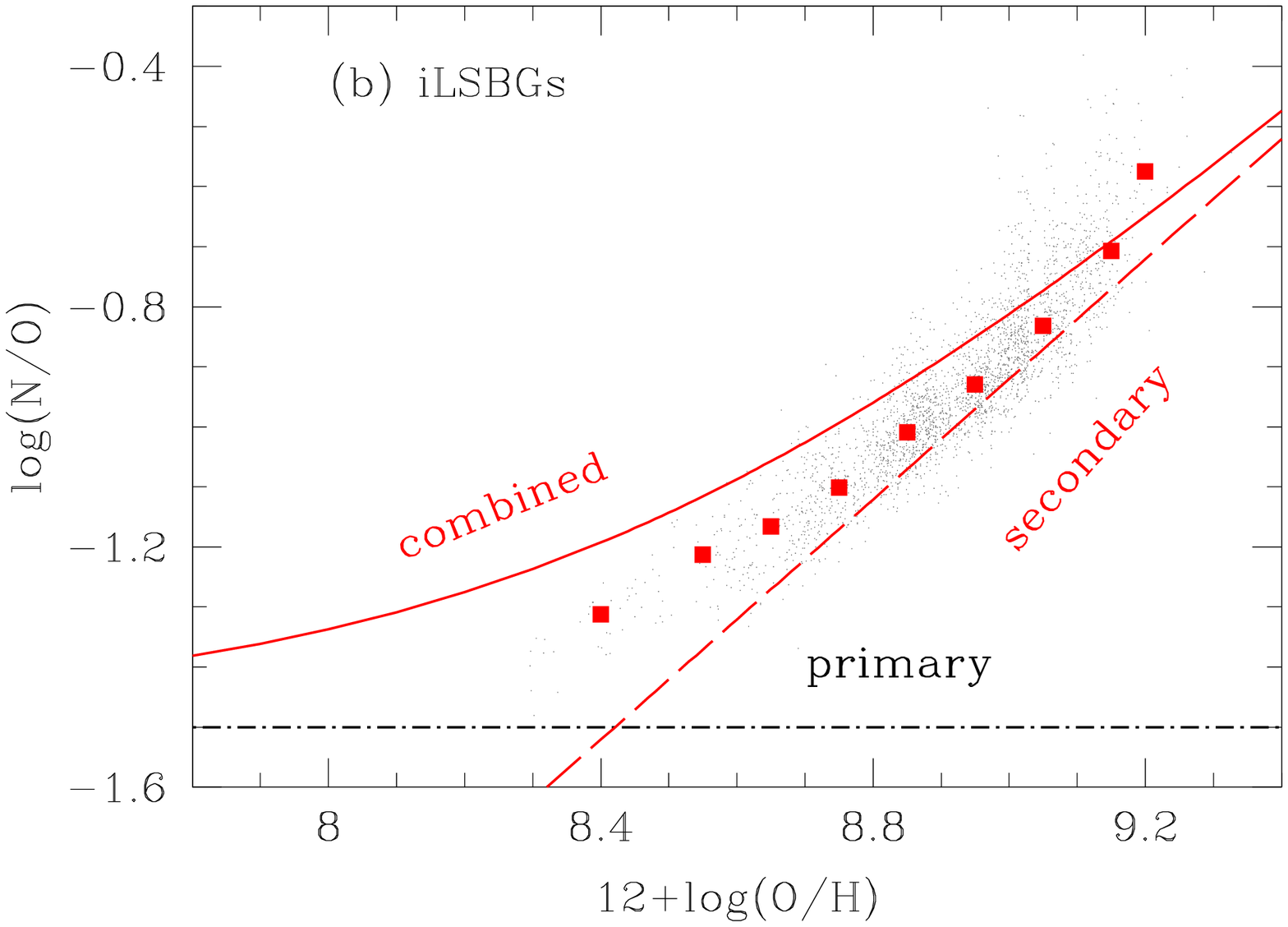} \\
\vspace{-1.8cm}
\includegraphics [width=7.5cm, height=6.5cm]{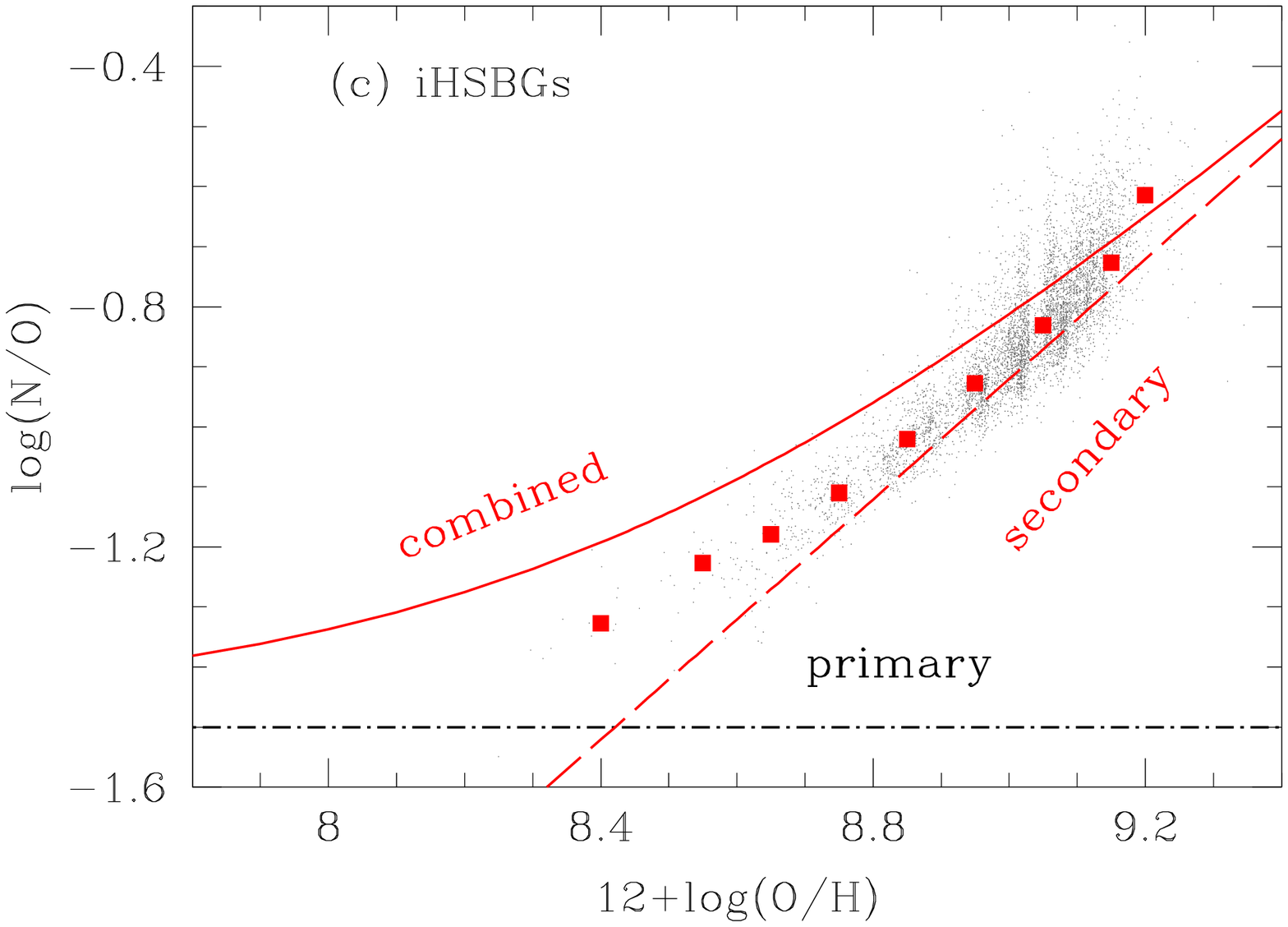} \\
\vspace{-1.8cm}
\includegraphics [width=7.5cm, height=6.5cm]{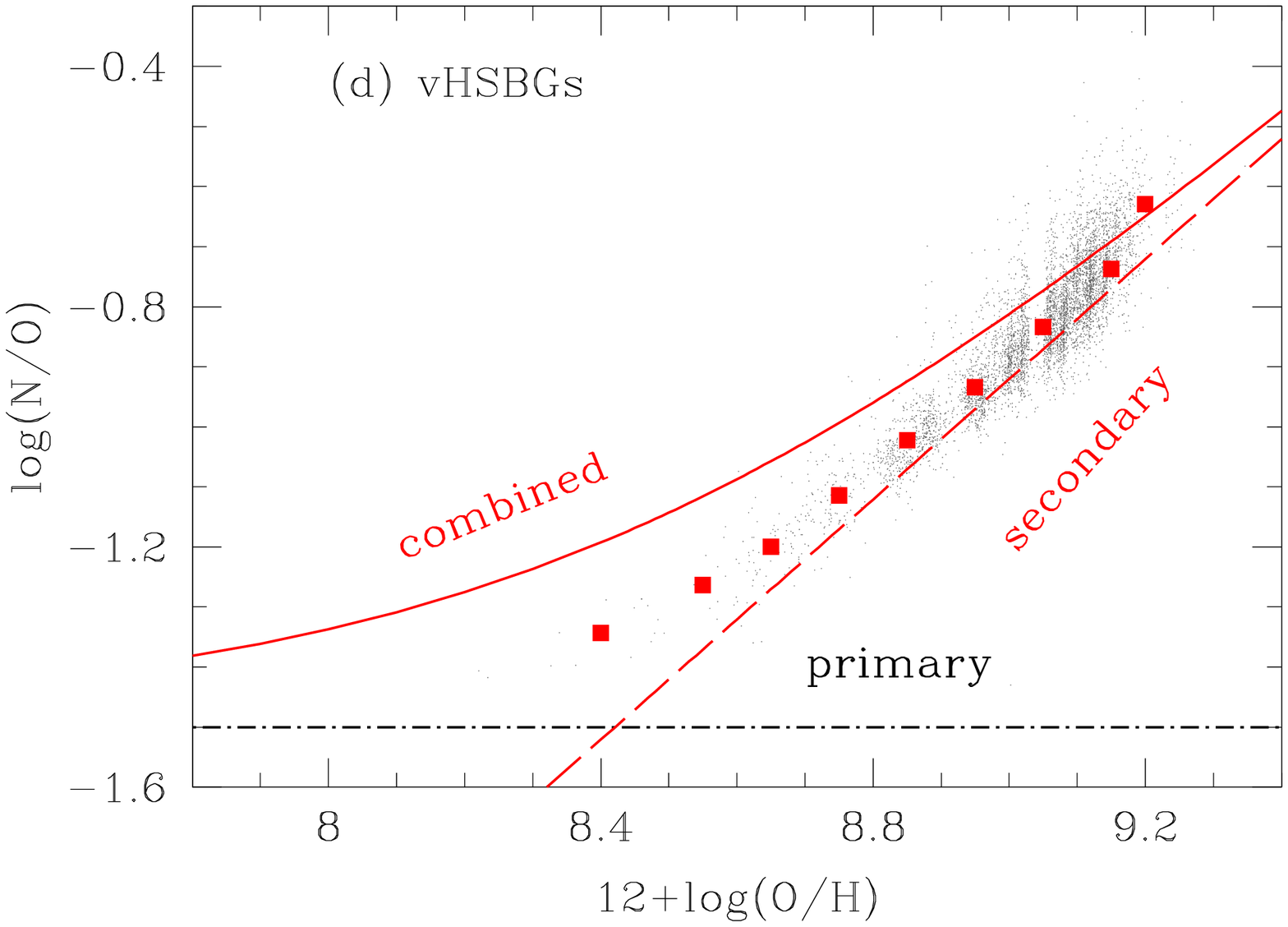} 
\vspace{-1.8cm}
\caption {The log(N/O) abundances of the star-forming sample galaxies
in four bins of $\mu_0(B)$:
 (a) vLSBGs, (b) iLSBGs, (c) iHSBGs, (d) vHSBGs. The ``primary" (the dot-dashed line),
``secondary" (the long-dashed line) components and the combination of these two
components (the solid line)  taken from Vila-Costas \& Edmunds (1993) 
have also been plotted. 
The big squares refer to the median values of log(N/O) 
with 0.1\,dex bins of 12+log(O/H). (Please see the on-line color version
for more details)}
\label{fig.no.oh}
\end{figure}

Figure~\ref{fig.no.oh}  presents the log(N/O) vs. 12+log(O/H) relations for our
sample galaxies in four bins of $\mu_0(B)$. 
It shows that
the vLSBGs have much less metal-rich and N/O-high
objects than the vHSBGs and vHSBGs. iLSBGs also have such less objects
than the two subgroups with higher surface brightness.
Thus the median values of log(N/O) are 
-1.05 for vLSBGs, 
-0.94 for iLSBGs, 
-0.85 for iHSBGs and 
-0.83 for vHSBGs (see Table~\ref{tab1}).
We also plot the different origin of ``primary" (the
dot-dashed line), ``secondary" (the long-dashed line) component and the 
combination of these two components (the solid line)  taken from
Vila-Costas \& Edmunds (1993) in this figure. 
It shows that the log(N/O) abundances
of the sample galaxies
are more consistent with the combination of the ``primary" and ``secondary"
components, but the $secondary$ component dominates. This result is similar
to the previous studies, e.g. Shields et al. (1991), Vila-Costas \& Edmunds
(1993), Contini et al. (2002), Kennicutt et al. (2003), Liang et al. (2006a),
and Mallery et al. (2007).
In Fig.~\ref{fig.no.oh},
the median values of log(N/O) of the four sub-sample galaxies 
are also given in each of the 0.1\,dex bins of 
12+log(O/H) as the big squares. 

\begin{figure} 
\centering
\includegraphics [width=7.5cm, height=6.5cm]{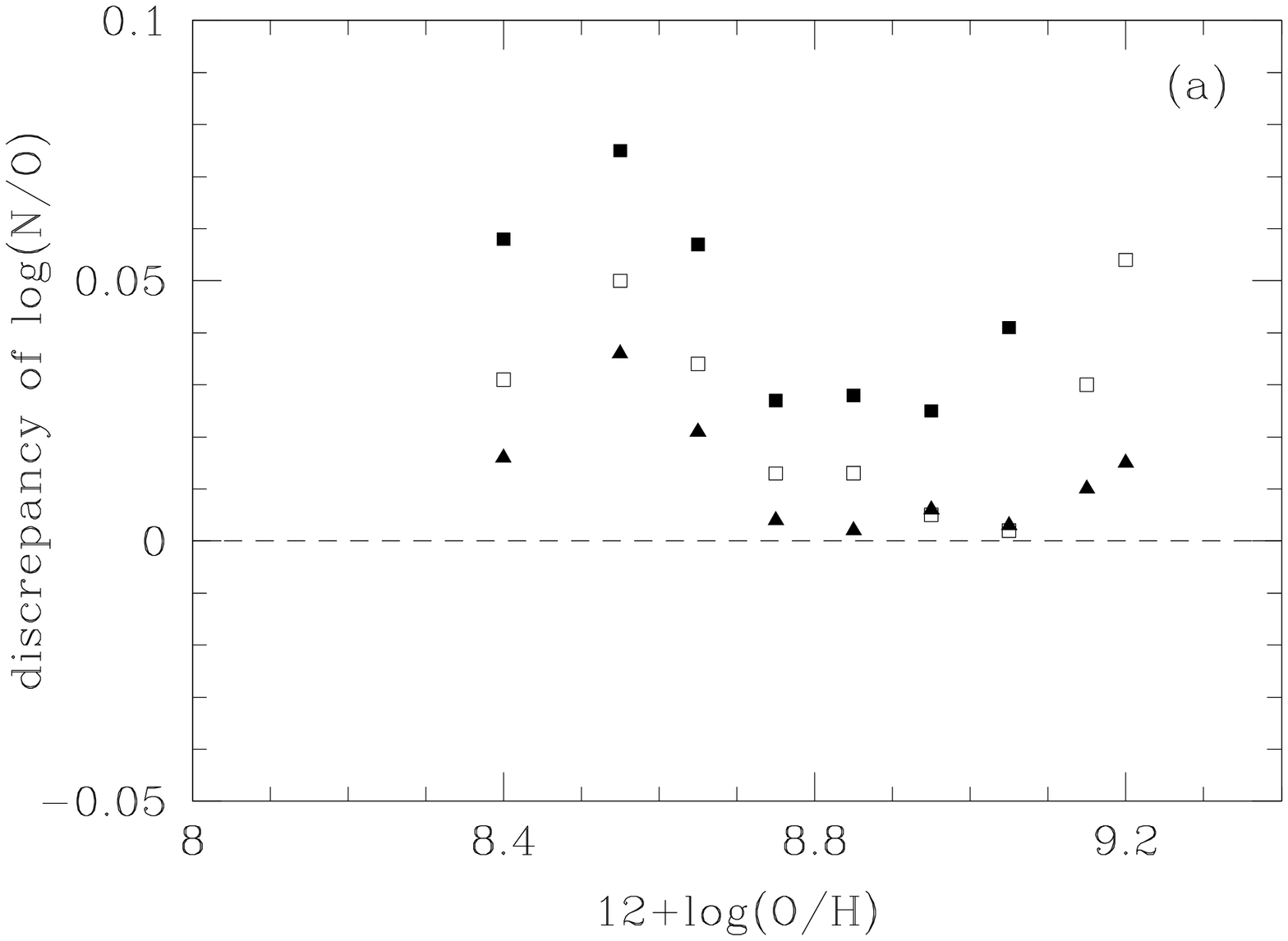} \\
\vspace{-1.8cm}
\includegraphics [width=7.5cm, height=6.5cm]{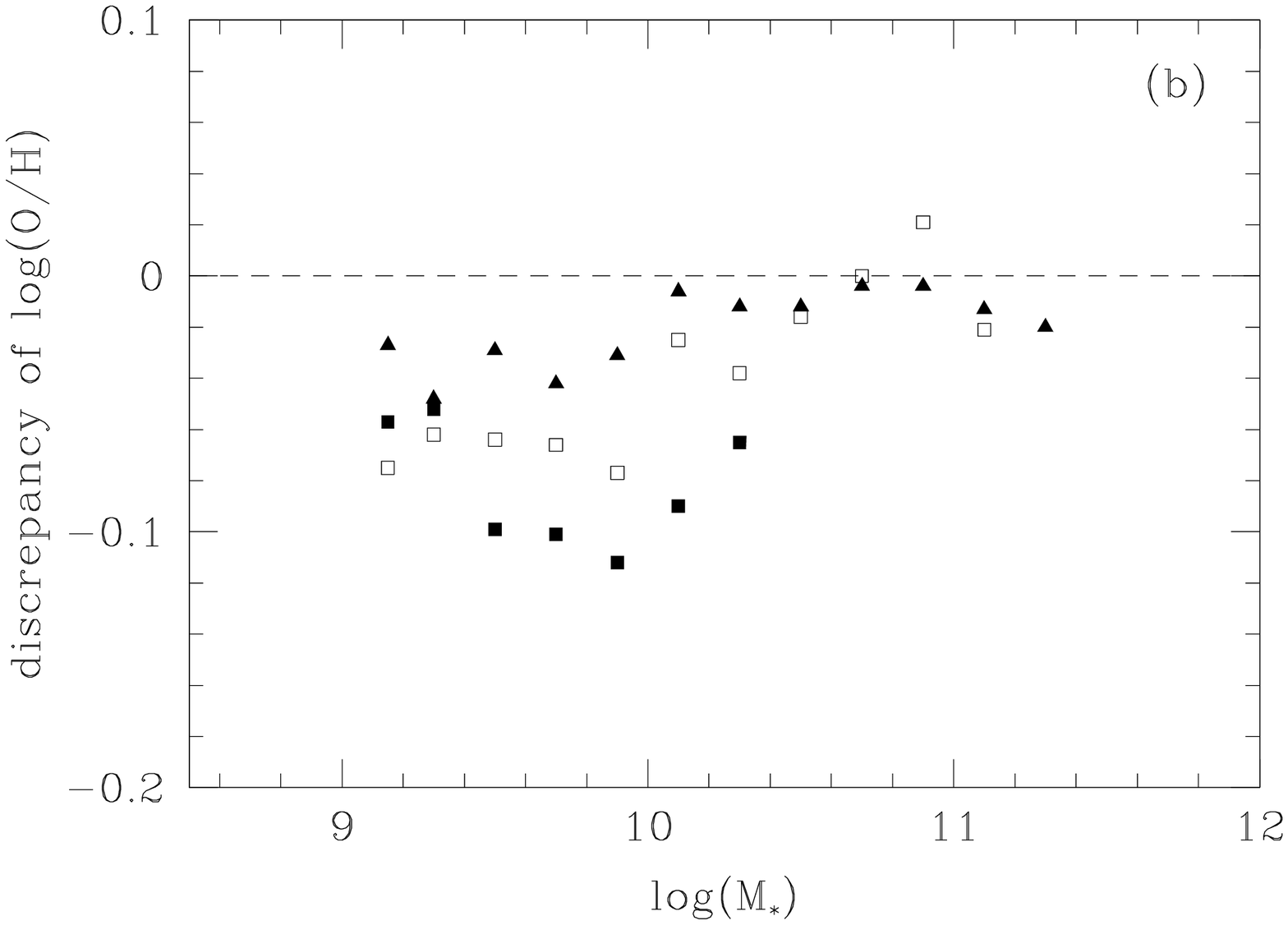} 
\vspace{-1.8cm}
\caption {The discrepancies among the median values
of the four subgroup galaxies in bins:
(a) the median values of log(N/O) vs. 12+log(O/H) as given in Fig.~\protect\ref{fig.no.oh};  
(b) the median values of 12+log(O/H) vs. stellar mass as given in Fig.~\protect\ref{fig.mass.oh}.
The filled squares refer to the difference of 
vLSBGs$-$vHSBGs, the open squares refer to the difference of 
iLSBGs$-$vHSBGs, and the filled triangles refer to 
the difference of iHSBGs$-$vHSBGs.}
\label{fig.dis.on.referee}
\end{figure}

To present the discrepancies 
among the median values of the four subgroup galaxies more clearly,
we plot their discrepancies in Fig.~\ref{fig.dis.on.referee}a,
where the filled squares are for 
vLSBGs$-$vHSBGs, the open squares are for 
iLSBGs$-$vHSBGs, and the filled triangles are for 
iHSBGs$-$vHSBGs. 
It shows that such discrepancies are more obvious, up to $\sim$0.074\,dex,
at low metallicity part. 
The reason could be 
related to different star formation rate (SFR) in the galaxies there as Molla et al. (2006) suggested.
Their Fig.5 clearly shows that 
the evolutionary track followed by a given region in the N/O-O/H 
plane depends strongly on the star formation history of the region:
strong bursting star formation histories would produce high oxygen
abundances soon and, hence, an early secondary behavior;
in contrast, a low and continuous SFR keeps the oxygen 
abundance low for a long time thus, a large quantity 
of the primary nitrogen may be ejected reproducing the flat slopes
in the N/O-O/H plane.
Combined these discussions with our Fig.~\ref{fig.no.oh} and Fig.~\ref{fig.dis.on.referee}a,
it suggests that the galaxies with lower 
surface brightness may have lower SFR, thus their N/O-O/H show more
primary nitrogen component at low metallicity region.

In the middle metallicity part, this discrepancy becomes much small, 
$\sim$0.03\,dex for vLSBGs$-$ vHSBGs, 
$\sim$0.015\,dex for iLSBGs$-$vHSBGs,
and almost no difference between iHSBGs and vHSBGs.
However, the discrepancies become obvious ($<$0.06\,dex) 
again when the metallicity becomes very high (12+log(O/H)$>$9.0).
The reason could be that possibly
there are more recently formed massive stars to produce
more oxygen elements, which will cause higher O/H but lower N/O since
nitrogen mainly comes from the intermediate-mass stars.
These differences in the median-value points suggest that
there might be some star formation history (SFH) variation with surface brightness, 
but to really test this we may need more data at low metallicity region,
and also in very high metallicity region.   
 However,
we should notice that the theoretical models 
 only predict very large differences in N/O vs. O/H with SFH at 
 metallicities below 12+log(O/H)=8.0. Thus, we really want 
 to go to low metallicity to test whether LSBGs and HSBGs have 
 had different SFHs. 

These results are consistent with what we found 
from the stellar population analyses through fitting spectral 
continua and absorption lines on these sample galaxies
as presented in Chen et al. (2010, in preparation), who found
that vHSBGs have larger fraction (5\%) of young population
than vLSBGs.  Also Gao et al. (2010) found that the derived ages
are 0.2 Gyr younger in HSBGs than in LSBGs generally through
fitting the FUV-to-NIR multiwavelength spectral energy distributions (SEDs)
of these galaxies
by using the PEGASE model (Fioc \& Rocca-Volmerange 1997). 
These results should be not much different from Mattsson et al. (2007), who 
found that the late-type LSB galaxies did not deviate from the general trend
in HSB galaxies and concluded
that LSB galaxies probably had the same age as their high surface brightness
counterparts,
although the global rate of star formation must be considerably lower in these
galaxies.

\section{ Relations among metallicity, stellar mass and surface
brightness}

We study the relations of 12+log(O/H) vs. stellar mass,
12+log(O/H) vs. $\mu_0$(B) and stellar mass vs. $\mu_0$(B) for our 
sample galaxies in this section.
 Due to the limits of stellar mass estimates, the available 
 samples in these relations are reduced to be
597 of vLSBGs,
2,609 of iLSBGs, 
5,486 of iHSBGs, and 
5,291 of vHSBGs.

\subsection{ Metallicity vs. stellar mass}

The stellar mass-metallicity relation (MZR) of galaxies is a 
fundamental relation to show
the evolutionary history and the present properties of galaxies. 
Generally, the metallicities and stellar masses 
of galaxies increase with their evolutionary processes.
Therefore, usually more massive
galaxies are more metal-rich 
(Tremonti et al. 2004; Liang et al. 2007; Kewley \& Ellison 2008 and the
references therein). Besides the local galaxies their MZR have been
obtained, 
a sample of galaxies at intermediate redshift ($0.4<z<1$) have also
been obtained their MZR based on the observed data from large telescope,
such as VLT and Keck (Liang et al. 2006b; Rodrigues et al. 2008),
even for the galaxies at high redshift ($z\sim 2.3$, Erb et al. 2006). 
 
\begin{figure} 
\centering
\includegraphics [width=7.5cm, height=6.5cm]{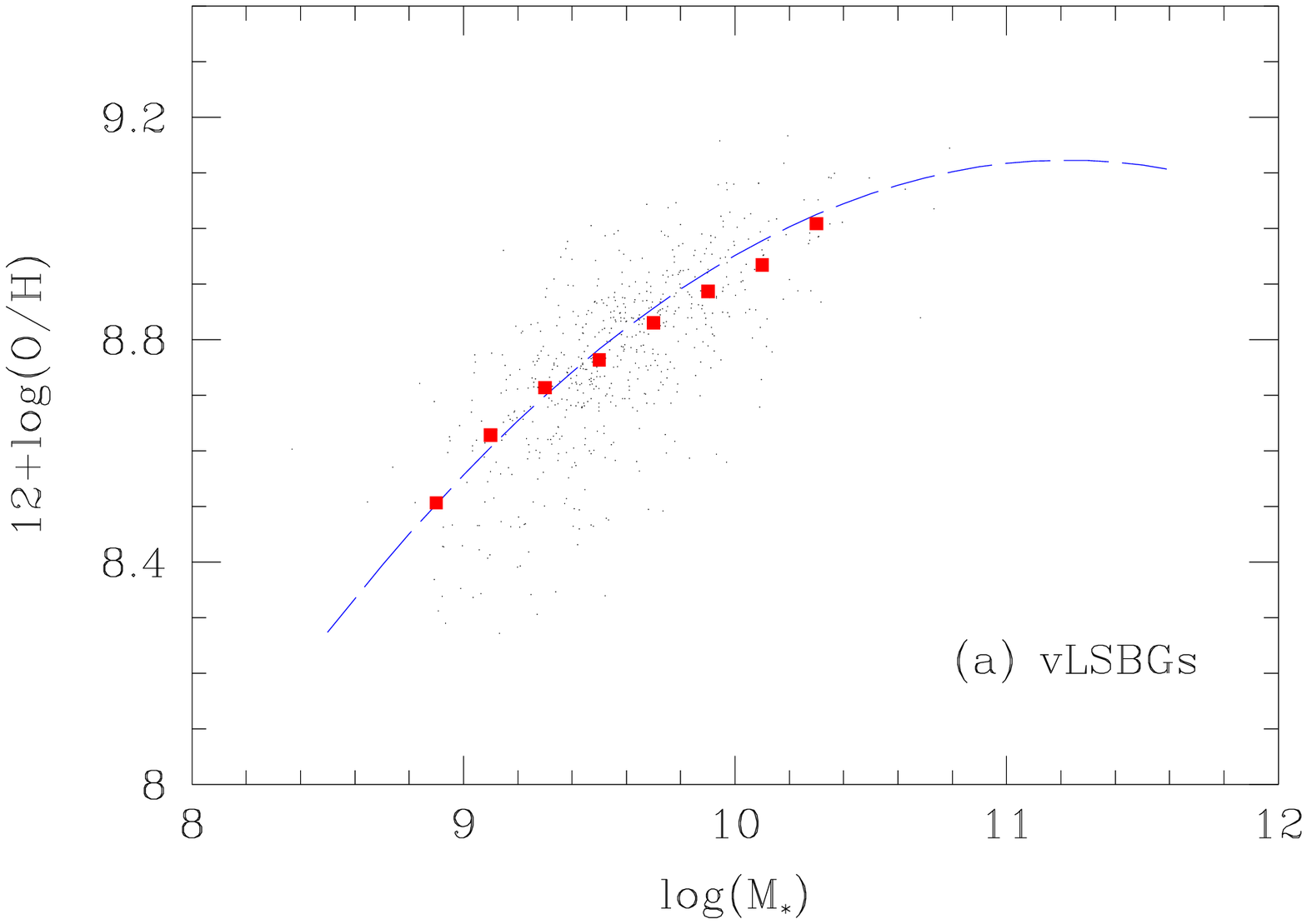} \\
\vspace{-1.8cm}
\includegraphics [width=7.5cm, height=6.5cm]{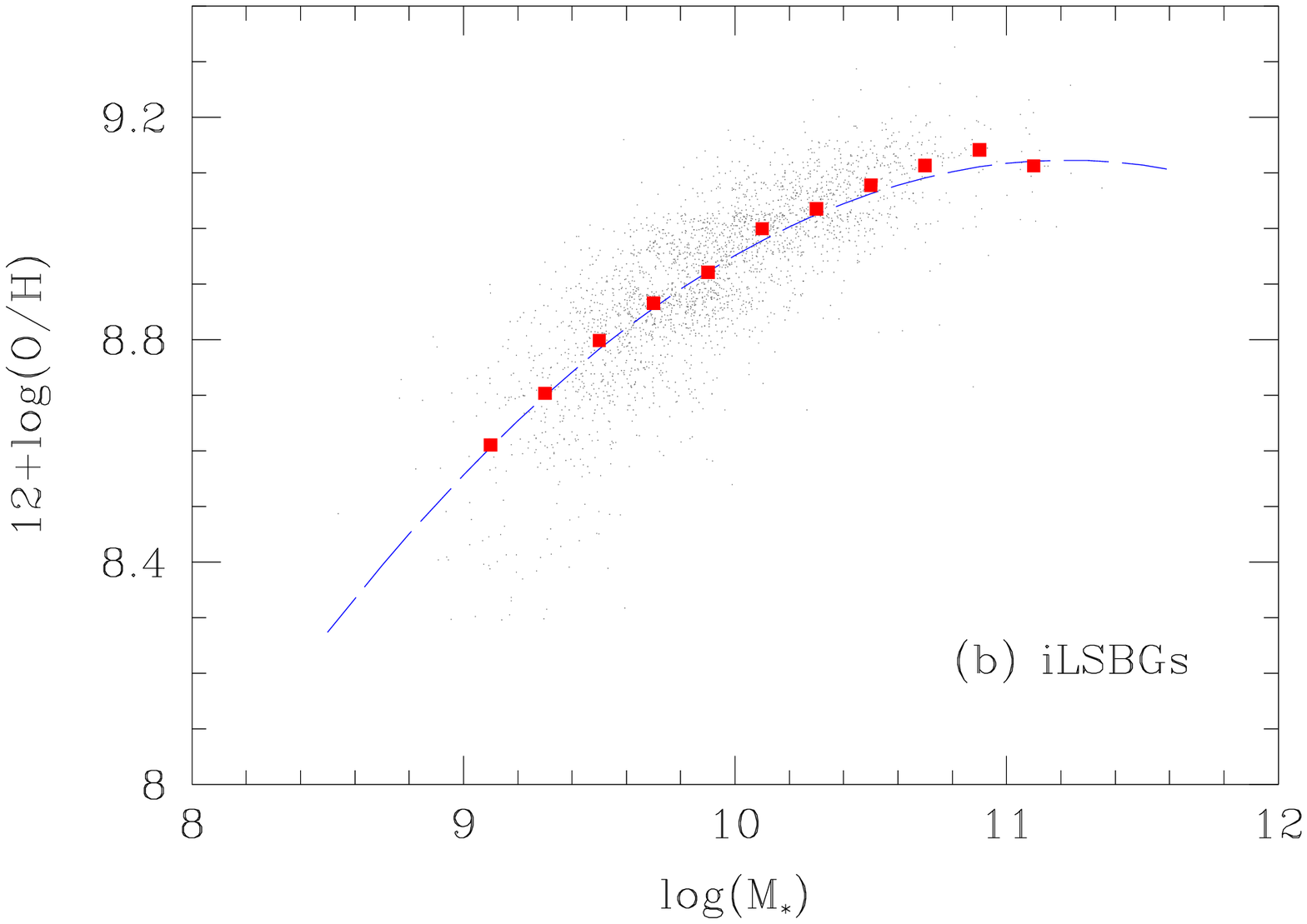} \\
\vspace{-1.8cm}
\includegraphics [width=7.5cm, height=6.5cm]{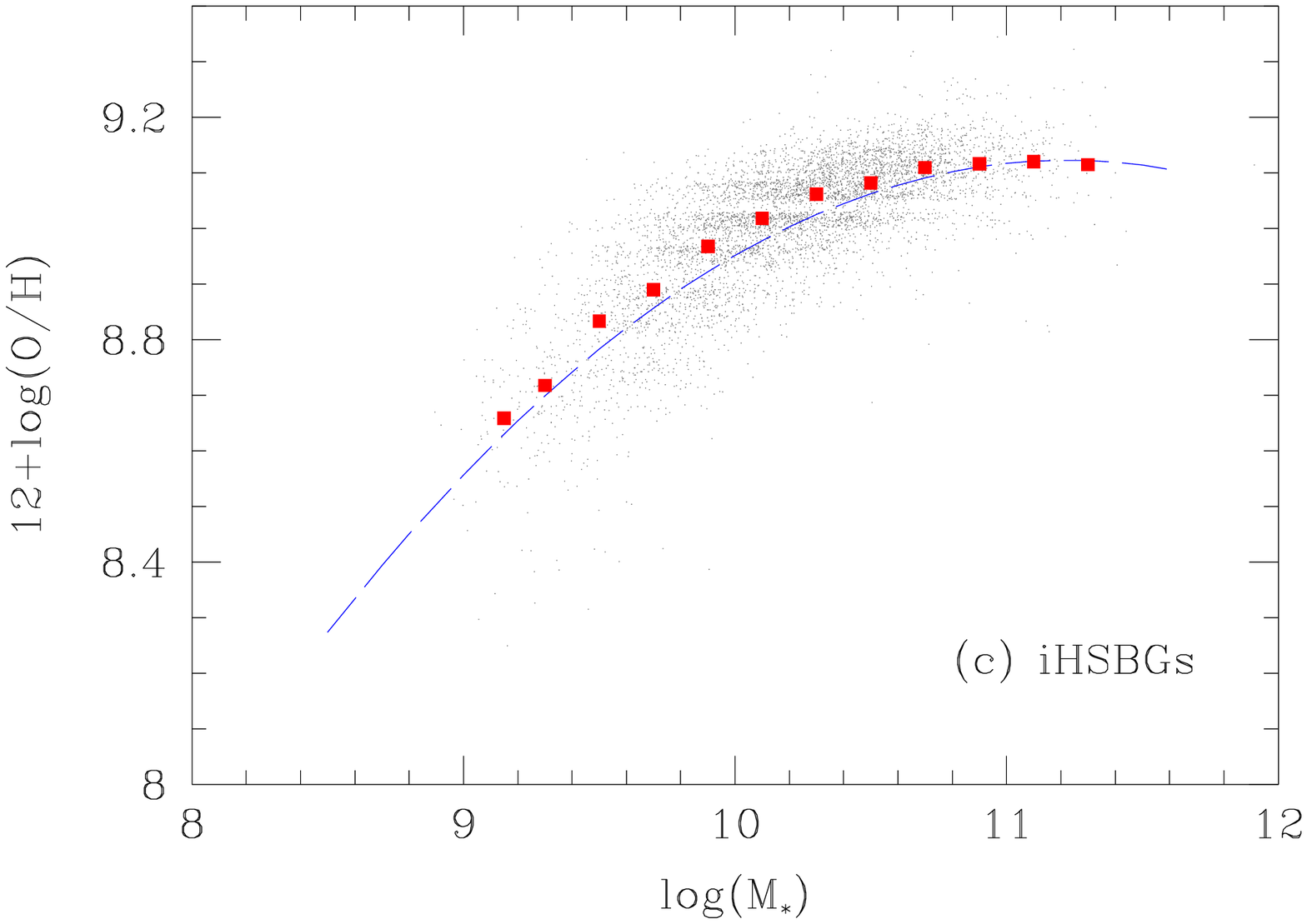} \\
\vspace{-1.8cm}
\includegraphics [width=7.5cm, height=6.5cm]{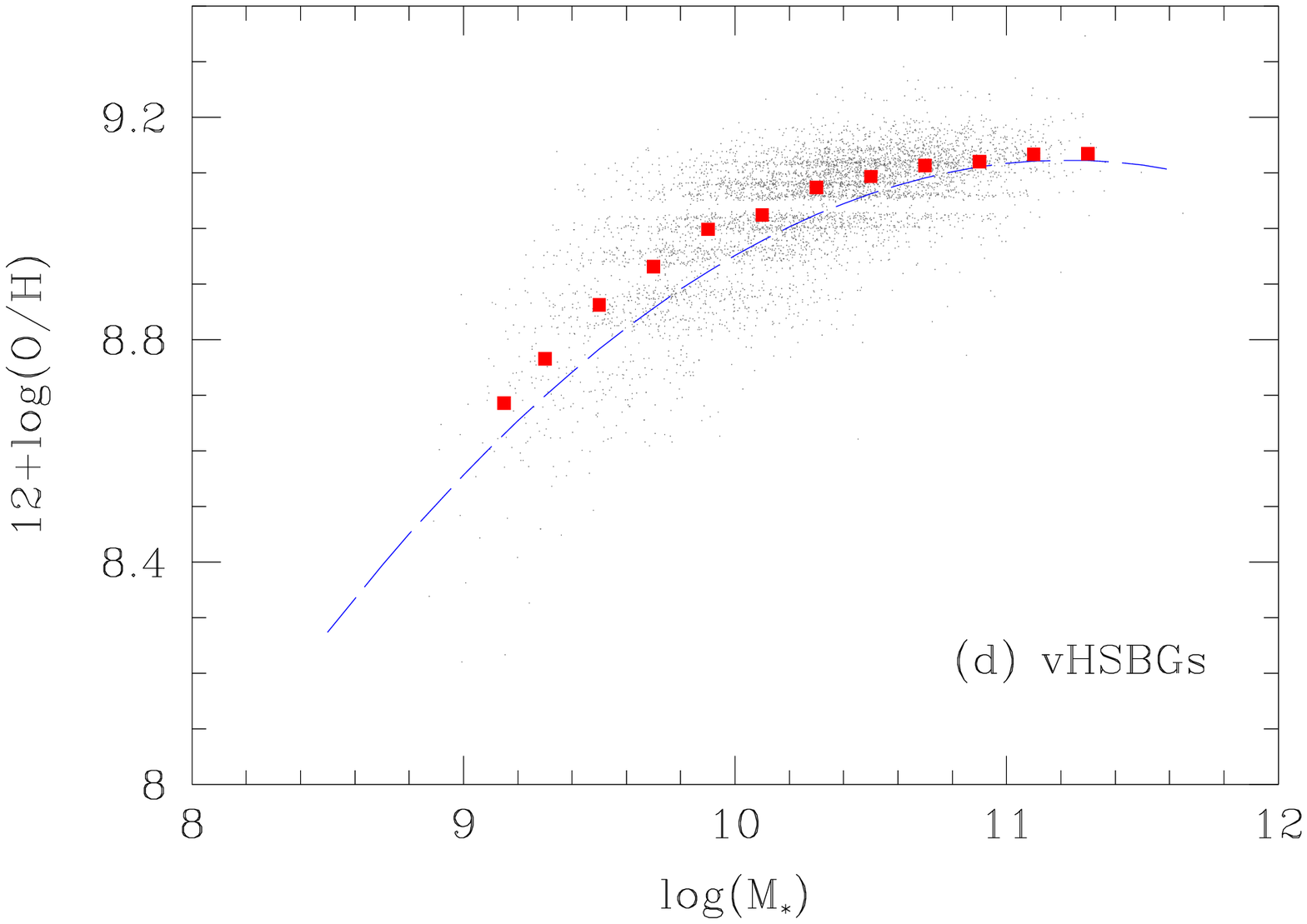} 
\vspace{-1.8cm}
\caption {The relations of stellar mass and 12+log(O/H) for the
star-forming sample galaxies in four bins of $\mu_0(B)$:
(a) vLSBGs,
(b) iLSBGs,
(c) iHSBGs,
(d) vHSBGs. 
The line is taken from Liang et al. (2007) for the 
SDSS-DR4 star-forming galaxies.
The big squares refer to the median values of 12+log(O/H) 
with 0.2\,dex bins of log(M$_*$). (Please see the on-line color version
for more details)
}
\label{fig.mass.oh}
\end{figure}

 Fig.~\ref{fig.mass.oh} presents the MZR for our sample galaxies in four
$\mu_0(B)$ bins. The median values of log(M$_*$) of the four subgroup
galaxies are 9.55, 9.91, 10.21 and 10.29 for 
vLSBGs, iLSBGs, iHSBGs and vHSBGs, respectively (see Table~\ref{tab1}).
 It shows that the vLSBGs include less samples of metal-rich and
 massive galaxies than the other three subgroups. 
 The iLSBGs have also less such objects than the iHSBGs and vHSBGs.
 Our samples are fall in the regions of the typical 
 SDSS star-forming galaxies 
 (see the dashed line in  Fig.~\ref{fig.mass.oh}, and Liang et al. 2007), 
 but with slight difference as discussed below and in Sect.\,4.3.
 
The median values of 12+log(O/H) with 0.2\,dex bins of stellar mass
are also given as the big squares in Fig.~\ref{fig.mass.oh} 
for the four subgroup galaxies.
Comparing these median values with the dashed line, 
HSBGs have slightly higher 12+log(O/H) ($\sim$0.035\,dex)  
than the normal galaxies at given stellar mass,
LSBGs have quite similar metallicities to them
except the vLSBGs with larger stellar masses 
have 0.016-0.039\,dex lower 12+log(O/H)
than the normal galaxies. 

To show more clearly the differences among the median values of
the four subgroup galaxies, in Fig.~\ref{fig.dis.on.referee}b we 
plot the discrepancies of vLSBGs$-$vHSBGs
as filled squares, the discrepancies of 
iLSBGs$-$vHSBGs as open squares, 
and the discrepancies of iHSBGs$-$vHSBGs as filled triangles.
It shows that vHSBGs have higher 12+log(O/H) abundances
than vLSBGs for lower mass galaxies, 0.06-0.11\,dex up to log(M$_*$)$\sim$10.3 
with the peak around log(M$_*$)$\sim$9.9.
The corresponding overabundances are 0.06-0.08\,dex for iLSBGs, and $\sim$0.04\,dex for iHSBGs. 
In more massive region, there is no much difference 
for their metallicities among iLSBGs, iHSBGs and vHSBGs. 
These results show that the MZR of galaxies have some differences
following their surface brightness.

In these 12+log(O/H) vs. log(M$_*$) relations of the four subgroup galaxies,
we also calculate  
the scatter of the data comparing with the median values in mass bins.
The results are about 0.12 for vLSBGs, 0.09 for iLSBGs, 0.08 for both
iHSBGs and vHSBGs.

\subsection{ Central surface brightness vs. metallicity}
 
 Figure~\ref{fig.mu.oh.mass}a shows the relations of 12+log(O/H) vs. $\mu_0(B)$ 
 for our sample galaxies in four subgroups. The median values
 in each of the bins of 0.2 in $\mu_0(B)$ are also given (the big
 squares), as well as the three vertical long-dashed lines at
 $\mu_0(B)$=21.25, 22.0 and 22.75 mag arcsec$^{-2}$ 
 to mark the ranges of $\mu_0(B)$
 for the four subgroups. A general varying trend shows that,
 for the vLSBGs and iLSBGs, the galaxies with lower surface
 brightness values have lower metallicities. The iHSBGs also following this trend
 slightly. The vHSBGs do not show such trend.
 
 McGaugh (1994) did not find direct correlation between 
 metallicities and surface brightnesses for their small sample.
 Since our sample is much larger, our results could be the first time
 to show such correlation between 12+log(O/H) and $\mu_0(B)$ for 
 LSBGs although the scatter is obvious.
 However, Bell \& de Jong (2000) has investigated the relation between
 the average metallicities inferred from the colors of each galaxy annulus
 and the average K-band surface brightness in that annulus (their Fig.7b).
 They found strong, statistically significant correlations between
 local metallicity and the K-band surface brightness.
 The LSBGs with lower surface brightness may 
 have less star forming and/or evolve more slowly than those with higher
 surface brightness.
 
In these 12+log(O/H) vs. $\mu_0$(B) relations,
the scatter of the data comparing with the median values in $\mu_0$(B) bins 
are about 0.15 for both vLSBGs and iLSBGs, 0.13 for 
iHSBGs, and 0.12 for vHSBGs. These scatters are larger than 
those in the 12+log(O/H) vs. log(M$_*$) relations.

\subsection{ Central surface brightness vs. stellar mass}

\begin{figure} 
\centering
\includegraphics [width=7.5cm, height=6.5cm]{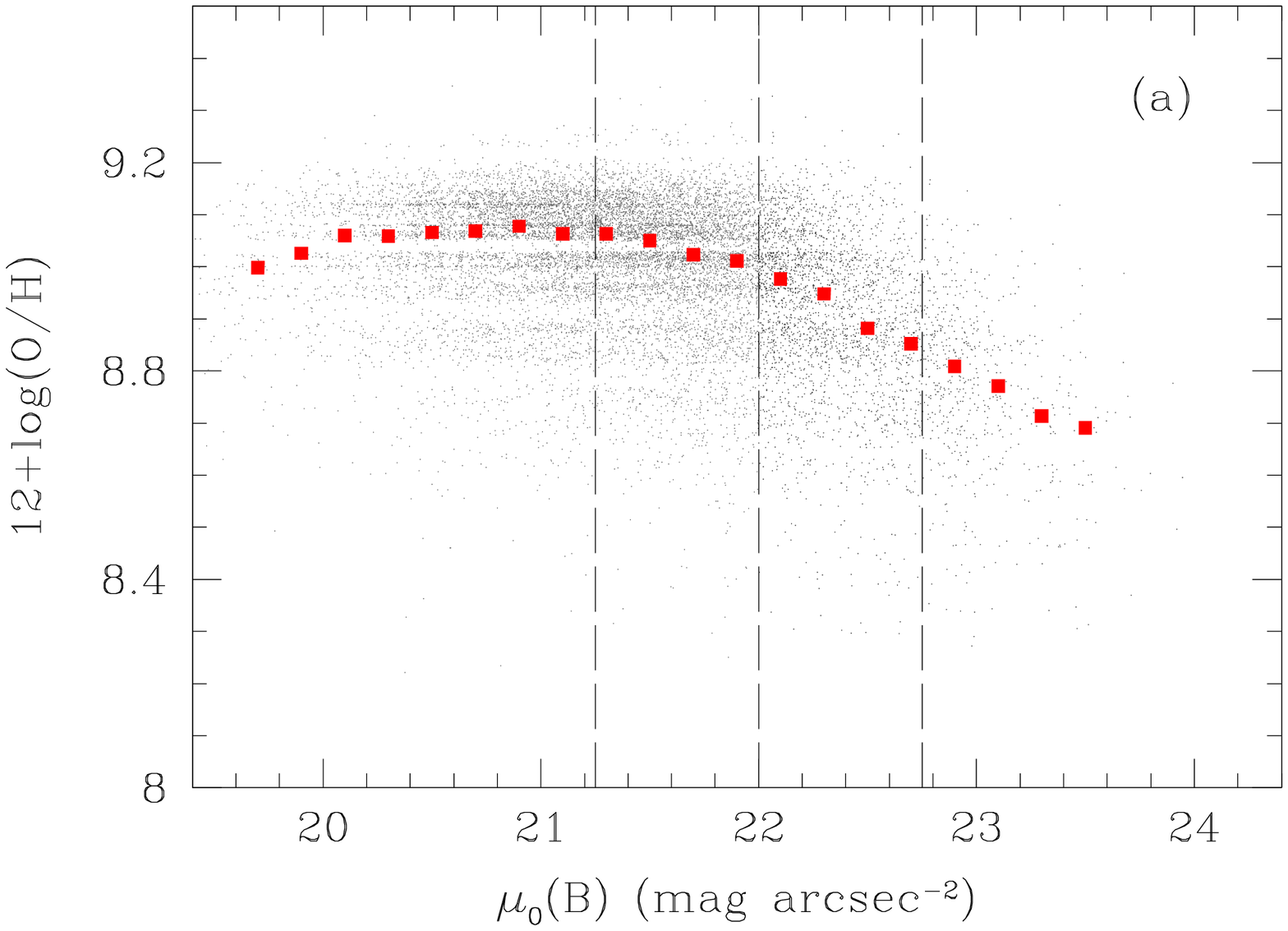} \\
\vspace{-1.4cm}
\includegraphics [width=7.5cm, height=6.5cm]{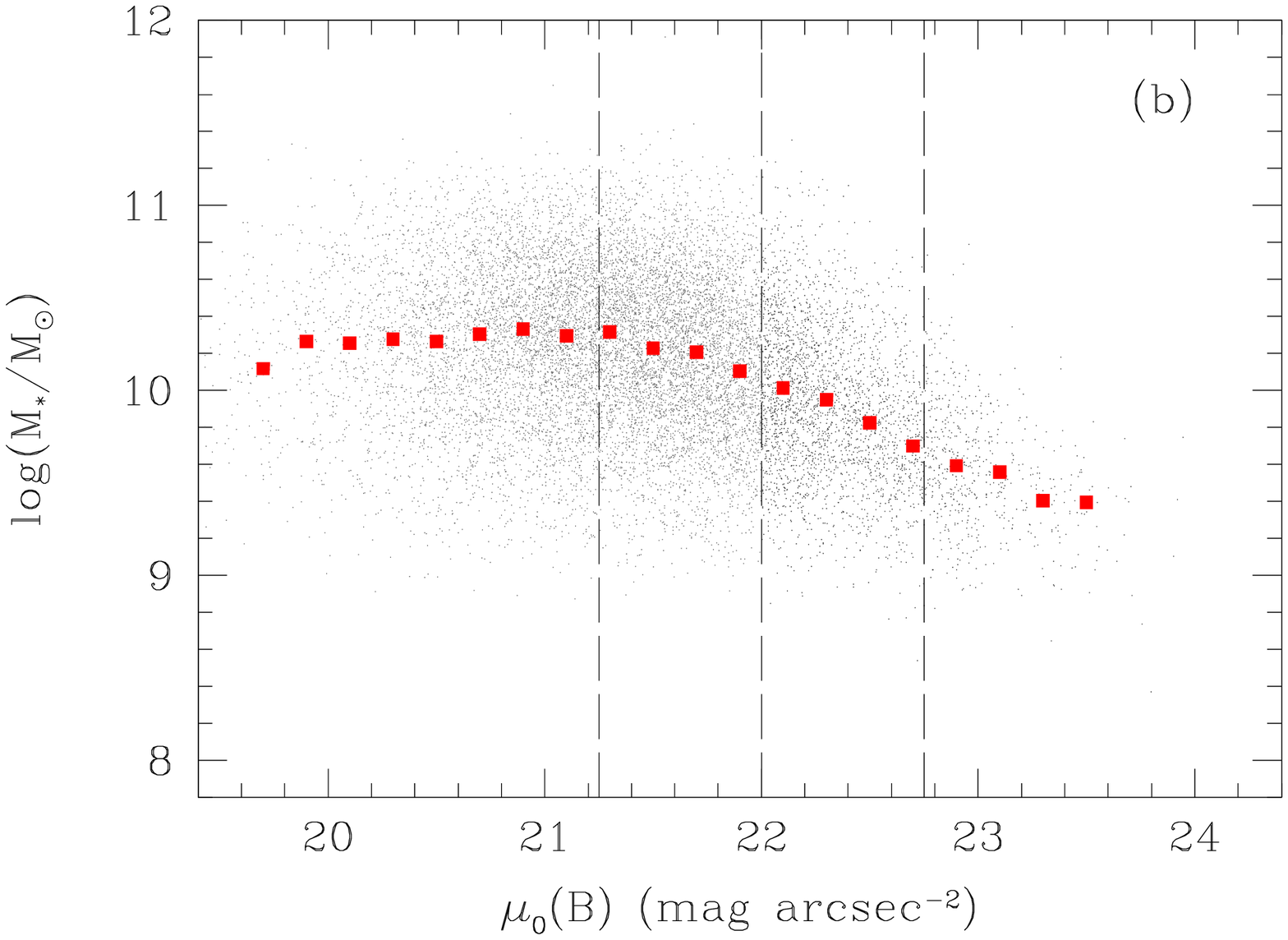} \\
\vspace{-1.4cm}
\includegraphics [width=7.5cm, height=6.5cm]{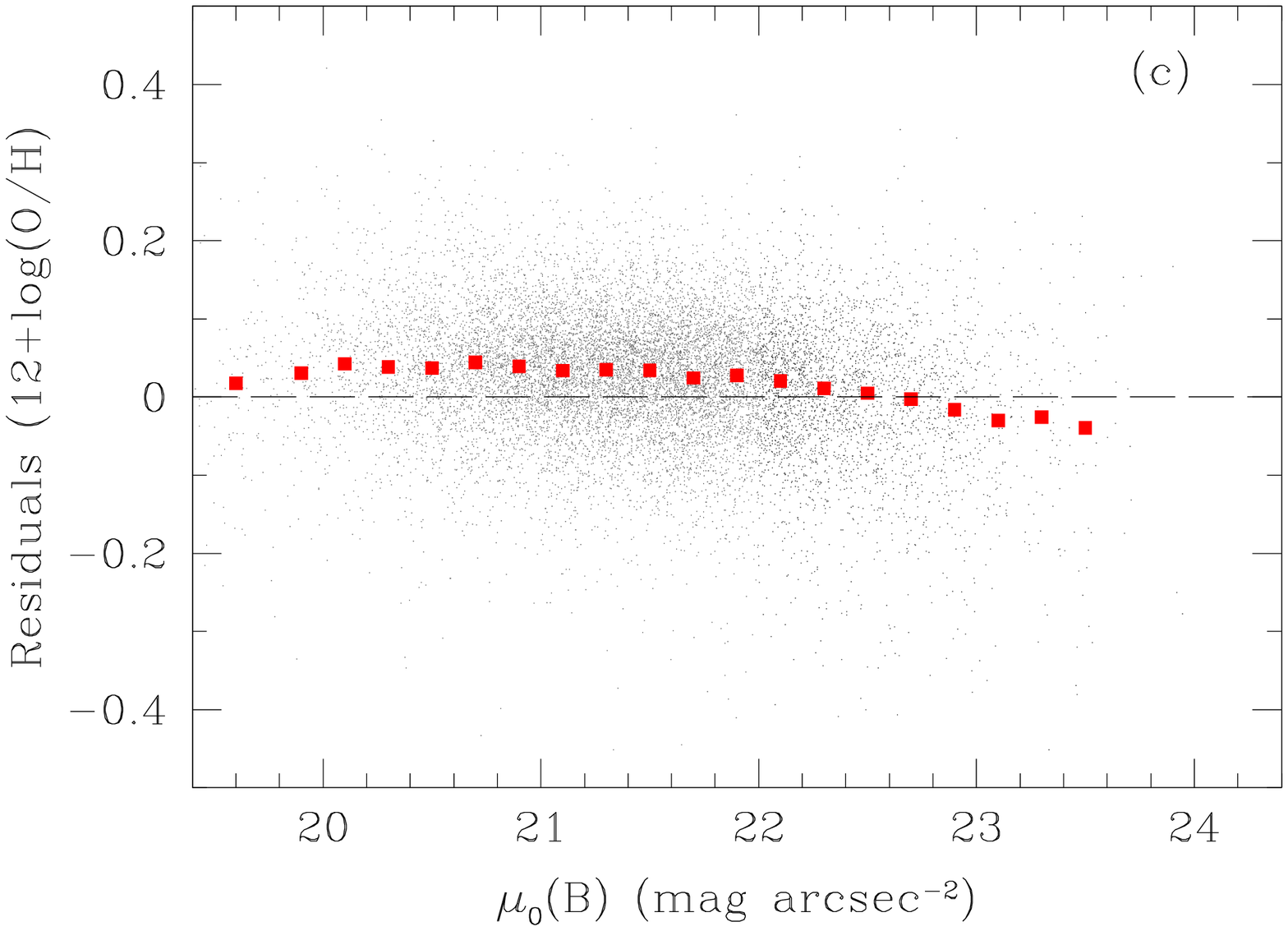} 
\vspace{-1.8cm}
\caption {The relations among 12+log(O/H), stellar masses and  $\mu_0(B)$ for
the star-forming sample galaxies:
(a) 12+log(O/H) vs. $\mu_0(B)$, 
(b) stellar mass vs. $\mu_0(B)$,  
(c) the residuals between the measured 12+log(O/H) and the ones calculated from the calibration 
with stellar mass (the dashed line in Fig.~\protect\ref{fig.mass.oh}, taken from Liang et al. 2007), 
which is to remove the stellar mass effects.
The median values in each of the bins of 0.2
in $\mu_0(B)$ are also given as the big squares, and the three
vertical long-dashed lines at 22.75, 22.0 and 21.25 mark the
ranges of $\mu_0(B)$ for the four subgroups of vLSBGs, iLSBGs,
iHSBGs and vHSBGs.}
\label{fig.mu.oh.mass}
\end{figure}

 We further obtain the relations of 
 $\mu_0$(B) vs. stellar masses for our sample galaxies, which are
 given in Fig.~\ref{fig.mu.oh.mass}b.
 Here the scatter is larger than 
 in Fig.~\ref{fig.mu.oh.mass}a.
 Similarly, it shows that, for the vLSBGs and iLSBGs,
 the galaxies with lower surface brightness have
 smaller stellar mass generally. The iHSBGs also follow this trend, but
 the vHSBGs do not. 
 Combining this relation with the relation of $\mu_0$(B) vs. 12+log(O/H), 
 these are consistent with the correlation between 
 stellar masses and metallicities of the sample galaxies.

 However, the trend of 12+log(O/H) following  
surface brightness for the vLSBGs and iLSBGs as shown in Fig.~\ref{fig.mu.oh.mass}a
could be linked with stellar masses. Therefore, in  Fig.~\ref{fig.mu.oh.mass}c
we plot the residuals of the measured 12+log(O/H) and the calculated ones
using the calibration formula from stellar mass derived 
by Liang et al. (2007) from the SDSS-DR4 star-forming galaxies,
i.e., the dashed line in our Fig.~\ref{fig.mass.oh}. 

The big squares in Fig.~\ref{fig.mu.oh.mass}c show the median values
of the metallicity residuals following surface brightness with bins
of 0.2. Comparing with vHSBGs, they show about 
0.035\,dex higher 12+log(O/H) for iHSBGs, 
quite similar metallicity for iLSBGs and about 0.027\,dex 
(0.016 to 0.039\,dex following decreasing surface brightness) lower
12+log(O/H) for vLSBGs.
These discrepancies following surface brightness
may be important, though
not the only thing, 
for 
the scatters of the data in relations
of 12+log(O/H) vs. stellar mass 
and 12+log(O/H) vs. surface brightness.

Moreover, these results are just 
consistent well with those shown in
Fig.~\ref{fig.mass.oh} (comparing the median-value points with the dashed line). 
These very small residuals are lower than what found by 
Tremonti et al. (2004, their Fig.7) with stellar surface density 
and Ellison et al. (2008, their Fig.2) with dividing by r-band half light radius.
The reason could be also related to our sample galaxies 
having very small fracDev$_r$ and being nearly face-on.
 It might also be the case that a
   galaxy's metallicity is more closely linked with its surface mass
   density than its B-band surface brightness. $\mu_0$(B) is likely very
   sensitive to any very recent perturbation in the SFH whereas
   metallicity will be more closely related to the time-integrated SFH.
In addition, this weak dependence
can also explain the 
relations of dust extinction $A_V$ with surface brightness 
as shown in Fig.~\ref{fig.Av},
which is consistent with the Fig.12 in Liang et al. (2007) about the 
relations of $A_V$ following log$M_*$ for the star-forming galaxies.
Then the 0.3\,dex difference in oxygen abundances 
between vLSBGs and vHSBGs (Fig.~\ref{fig.his.oh}) could be also understood well.

\section{Discussions}

\subsection{The Aperture effects}

Here we discuss the fiber aperture effects on metallicity estimates
for our sample galaxies.
Tremonti et al. (2004) has discussed the weak effect of the 
3{$^{\prime \prime }$} aperture of the SDSS spectroscopy 
on estimated metallicities of the sample
galaxies with redshifts $0.03<z<0.25$. 
We have used redshift $0.04<z<0.25$ as one of the sample selection criteria
by following
Kewley et al. (2005), who 
recommended that, to get reliable metallicities, redshifts $z>0.04$ are
required for the SDSS galaxies to ensure a covering fraction $>$20\% of the
galaxy light.

\begin{figure} 
\centering
\includegraphics [width=6.5cm, height=5.5cm] {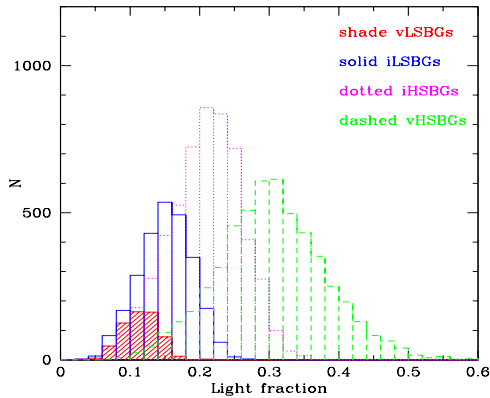} 
\caption {The histogram distributions of the light fraction 
(Eq.~(\protect\ref{apar}))
with bins of 0.01
for the sample galaxies in four subgroups. 
The median values of the light fractions in four bins of $\mu_0$(B) are 
0.12 for vLSBGs, 
0.15 for iLSBGs, 
0.21 for iHSBGs, 
0.30 for vHSBGs.
}
\label{fig.a.1.5}
\end{figure}

To check how much the light of the galaxy was covered by the fiber observation,
  one simple
   and accurate way is to compare the ``fiber" and
   ``petrosian" magnitudes of the SDSS galaxies.  
   The fiber mag is a measurement of
   the light going down the fiber and the petrosian mag is a good
   estimate of the total magnitude. Thus, we adopt the formula below
   to estimate how much light was covered by the fiber observations: 
\begin{equation}     
   light\_fraction = 10^{(-0.4*(fiber\_{mag} - petro\_{mag})_r)}.
\label{apar}
\end{equation}     
 Fig.~\ref{fig.a.1.5} shows the calculated light fractions for the four
 subsamples. 
It shows that, the light fractions of vLSBGs (597), iLSBGs (2,609), 
iHSBGs (5,486) and vHSBGs (5,291)
are about 0.12, 0.15, 0.21 and 0.30 (in median), respectively. 
Although the light fractions of vLSBGs and iLSBGs are lower than 
0.2 which was suggested by Kewley et al. (2005), 
we believe that this is not a serious problem
for estimating the metallicities of the sample galaxies 
since LSBGs may have no strong radial oxygen abundance gradients, which
has been found by de Blok \& van der Hulst (1998).

To further check whether 
the presence of a stellar bulge affects the
   relationships among stellar mass, metallicity and surface brightness
   of galaxies or not, 
we only select those samples having fracDev$_r$$\sim$0, which means that the
galaxy could be purely disk one.
Then we re-plot Fig.~\ref{fig.mu.oh.mass} and find  
they show very similar distributions to
the previous plots. 
Thus, we believe that the bulge light
won't affect much these corresponding relationships
for the sample galaxies.
This is consistent with what have been found by Ellison et al. (2008)
and Tremonti et al. (2004).  Ellison et al. (2008)
 showed that Bulge-to-total ratio has almost no impact
 on the mass-metallicity relation (their Fig 3) and Tremonti et
 al. (2004) showed that Concentration is also not important (their Fig. 7).
But we should still keep in mind that generally the central regions
have 
higher stellar surface density 
 (independently of the presence of a bulge) 
than outer regions, and more metal rich.

\subsection{Comparisons with previous studies on metallicities of LSBGs}

We find that  
our large sample of LSB disc galaxies selected from the SDSS
have no that low metallicities  
as the H~{\sc ii} regions in a small sample of LSBGs 
studied in literature.
The median 12+log(O/H) of
our vLSBGs (with $\mu_0(B)>$22.75 mag arcsec$^{-2}$),  
is 8.77, 
and the median 12+log(O/H) of the iLSBGs (with $\mu_0(B)$=22.75-22.0 mag arcsec$^{-2}$)  
is 8.94.
However, some researches have been obtained 
much lower oxygen abundances for the 
H~{\sc ii} regions in a small sample of LSBGs, i.e.,
8.06 to 8.20 of 12+log(O/H) (in median in their samples) 
in McGaugh (1994), Roennback \& Bergvall (1995),
de Blok \& van der Hulst (1998) and
Kuzio de Naray et al. (2004).
How to understand this discrepancy?

The most possible reason could be our sample galaxies are
more luminous than theirs indeed. 
We have selected the sample with
$M_B<-$18 mag which exclude some dwarf galaxies from 
the normal star-forming disc galaxies. Most of our sample 
galaxies are luminous with $M_B<-$19 mag.
Another reason could be that they (mostly)
adopted the $R_{23}$ calibration of lower branch, which surely
result in low oxygen abundances, while  
our oxygen abundances are those obtained by the MPA/JHU 
group as Tremonti et al. (2004) by using the Bayesian approach.
Moreover, the different photoionization models used
to estimate oxygen abundances could also cause some differences.
They used McGaugh (1991) which results 
in $\sim$0.1\,dex lower oxygen abundances than what used by Tremonti et al. (2004). 
In addition,
the metallicities from literature come from H~{\sc ii} regions which are often
   observed at large radii in the galaxy, whereas our measurements
  come from the central few kpc. If there are metallicity gradients in the galaxies
  then this will cause the SDSS metallicities to be a bit higher.  

In McGaugh (1994), 41
H~{\sc ii} regions in 22 LSBGs were obtained their O/H  abundances, which
are about 12+log(O/H)$\sim$8.06 (in median).  The sample
was selected from the lists of Schombert \& Bothun (1988) 
and Schombert et al. (1992), and from the UGC. All
these galaxies have central surface  brightness well below the
Freeman (1970) value of $\mu_0=21.65$ mag arcsec$^{-2}$, with the sample median
being 23.4 mag arcsec$^{-2}$.
Their Fig.10 shows that almost half of their samples are faint having
$M_B>-$18 mag, which should belong to the dwarfs, and
another half objects mostly have $-$20$<M_B<-$19 mag 
(the cosmological parameter $H_0$ has been corrected 
from 100 to 70 km s$^{-1}$ Mpc$^{-1}$). 
Even these brighter galaxies are comparable with part of our
sample galaxies in luminosity, 
McGaugh (1994) used the lower branch formula of $R_{23}$ for metallicity 
estimates, thus the oxygen abundances of their samples are still
lower than ours as discussed below. 

Another key point is which calibration (the lower branch or upper branch one) 
are used to estimate the 
oxygen abundances of the galaxies.
McGaugh (1994) used log([N~{\sc ii}]/[O~{\sc ii}])$<-$1.0 to judge that
their samples belong to the lower branch of metallicity, and then
used the lower branch $R_{23}$ formula (McGaugh 1991) to calculate
the oxygen abundances of their H~{\sc ii} regions.
Indeed their Fig.3 shows that
some (almost half) of their samples have log([N~{\sc ii}]/[O~{\sc ii}])
between $-$1.0 and $-$1.2. The latter value, $-$1.2, 
is a recent recommendation (Kewley
\& Ellison 2008) and is adopted here 
by us to separate the
upper and lower branches of metallicities.

\begin{figure} 
\centering
\includegraphics [width=4.1cm, height=4.0cm]{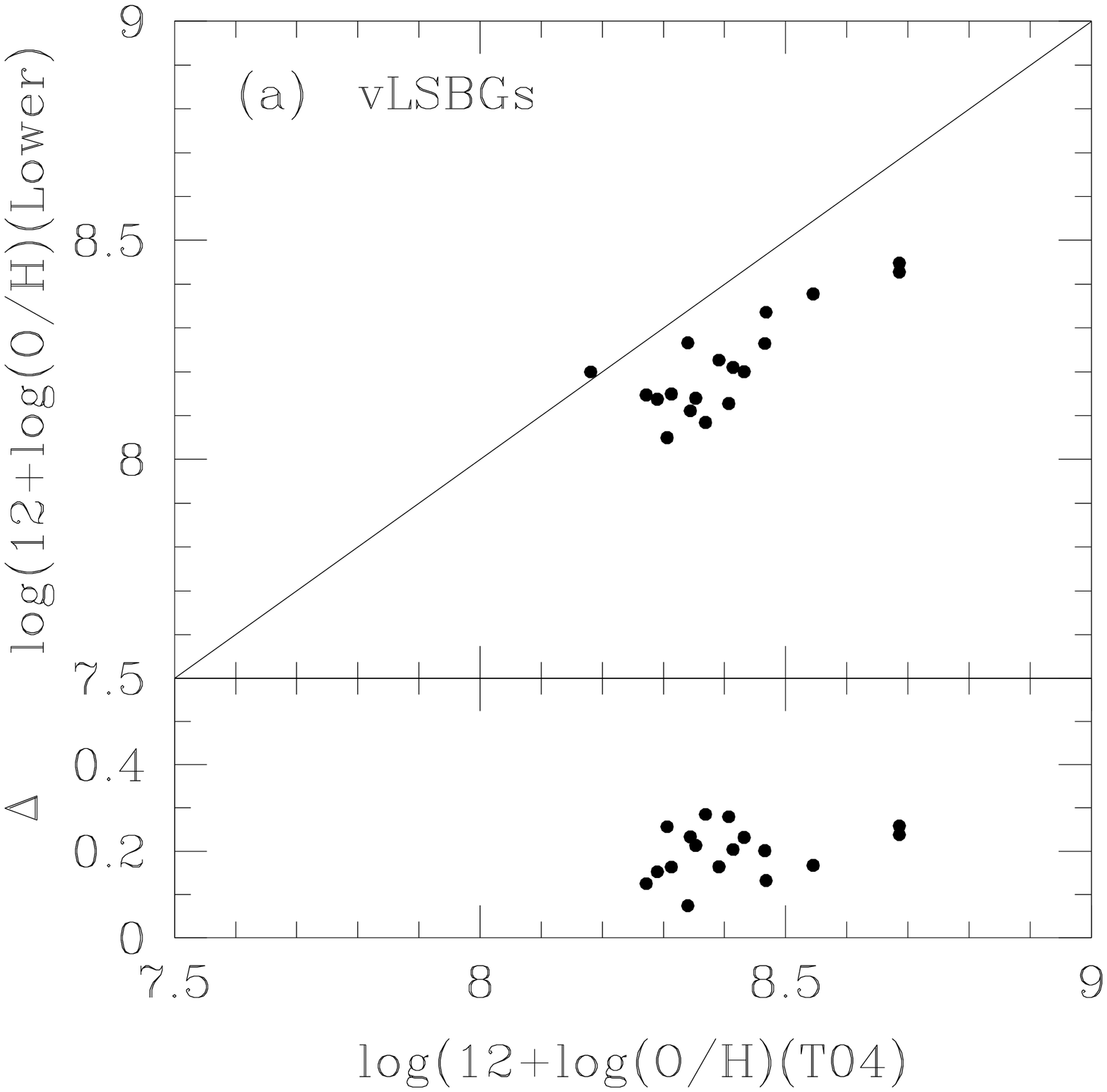} 
\includegraphics [width=4.1cm, height=4.0cm]{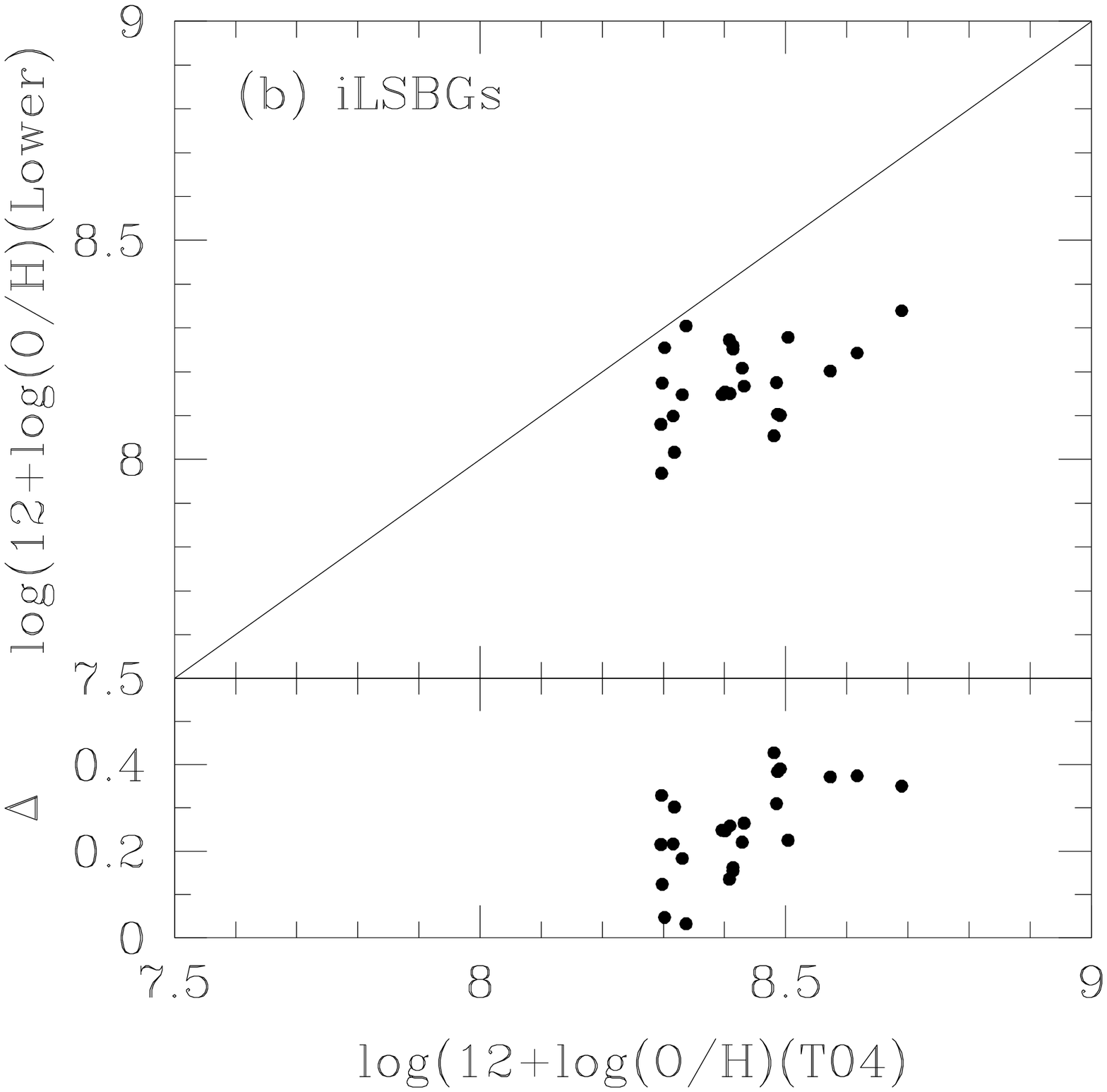}\\
\includegraphics [width=4.1cm, height=4.0cm]{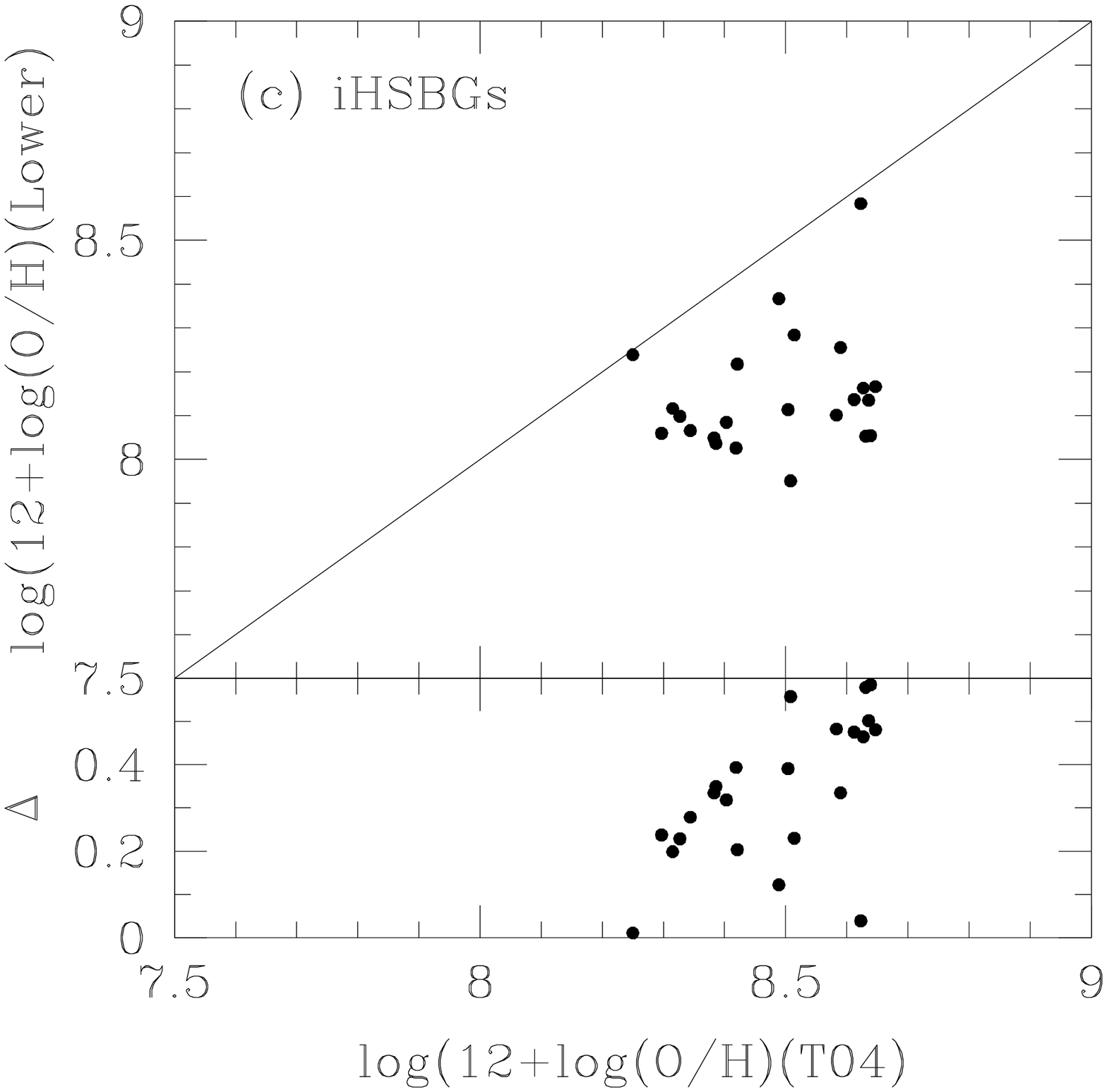} 
\includegraphics [width=4.1cm, height=4.0cm]{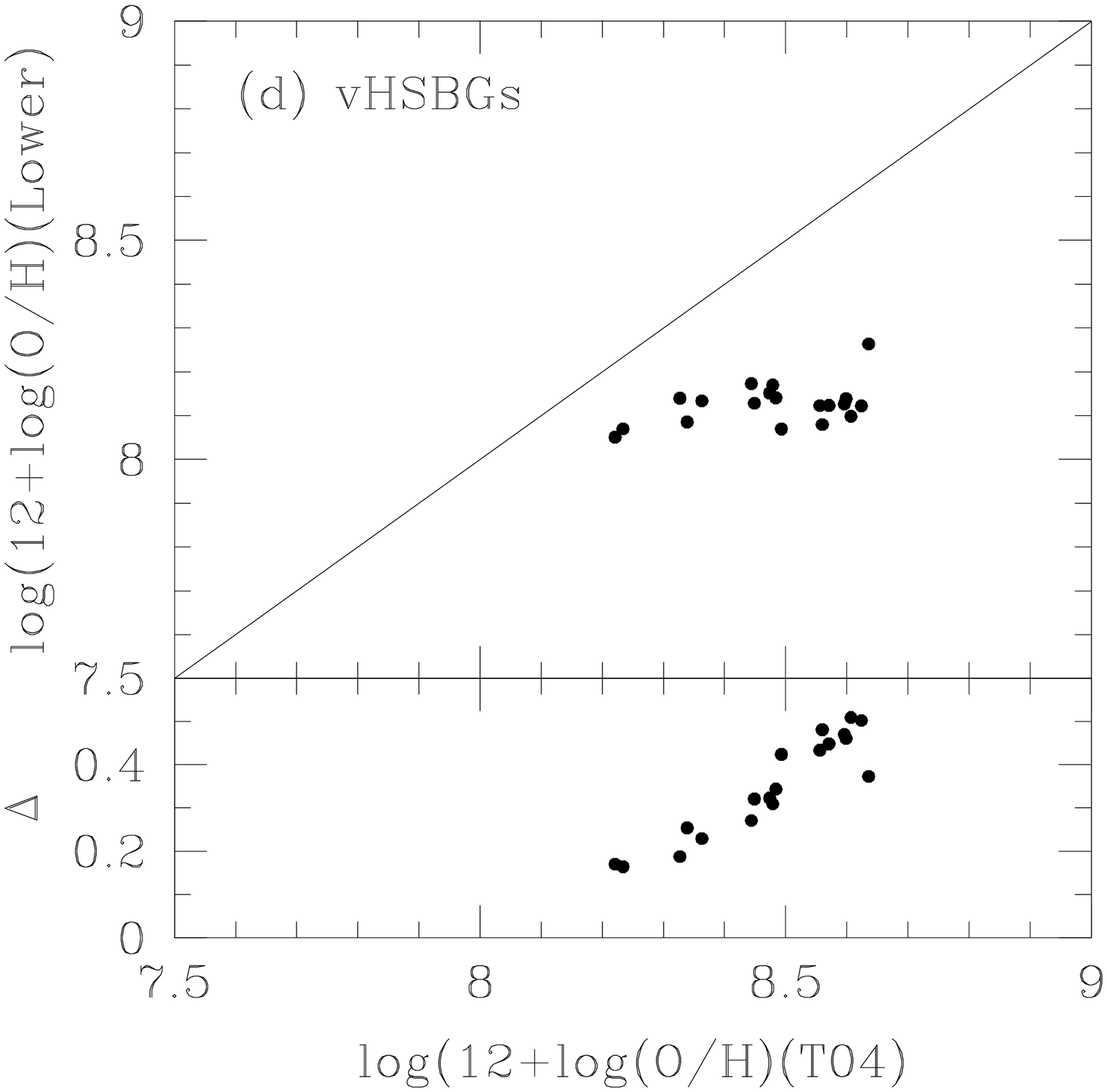}
\caption {Comparing the 12+log(O/H) abundances estimated from 
lower branch formula of $R_{23}$ (McGaugh 1991; Kobulnicky et al. 1999)
and the abundances from Tremonti et al. (2004) for the sample galaxies having 
$-$1.2$<$log([N~{\sc ii}]/[O~{\sc ii}])$<-$1.0:
 (a) vLSBGs, (b) iLSBGs, (c) iHSBGs, (d) vHSBGs.
The top panels show the direct comparisons, and the lower panels show the 
residuals.
}
\label{fig.com.oh.lower}
\end{figure}

To be sure the calibration method is another main reason (may be the most important one) 
for the difference between the abundances of McGaugh (1994) and ours,
we select those of our sample galaxies having $-$1.2$<$log([N~{\sc ii}]/[O~{\sc ii}])$<-$1.0
and re-calculate their oxygen abundances by using the same method as McGaugh (1994),
i.e., the lower branch $R_{23}$ formula
from McGaugh (1991). The corresponding analysis formula 
from Kobulnicky et al. (1999) is used here.
Then Fig.~\ref{fig.com.oh.lower} explains the differences quite well. 
It shows that the 12+log(O/H) of all these sample galaxies 
will become much lower, 0.2\,dex-0.6\,dex lower, than the original ones.
And now they are consistent with the estimates of McGaugh (1994). 

In addition, de Blok \& van der Hulst (1998) present measurements of the oxygen abundances
of 64 H~{\sc ii} regions in 12 LSBGs, and found their oxygen abundances are low,
in a range of 12+log(O/H)$\sim$ 7.83 to 8.84, 
and the median value is 8.20.
They confirm the results of McGaugh (1994) that LSBGs are metal-poor. 
However, most of their sample galaxies are the same ones as in 
McGaugh (1994). Also almost half of their samples have 
$-$1.2$<$log([N~{\sc ii}]/[O~{\sc ii}])$<-$1.0, but they use the lower branch $R_{23}$ formula
for them.
Kuzio de Naray et al. (2004) 
 combined their new
   abundance measurements with data from McGaugh (1994) and 
   de Block \& van der Hulst (1998) to
   produce average abundances for 18 galaxies.
They found that their galaxies to be
   significantly offset from the Tremonti's luminosity-metallicity
   and stellar mass-metallicity relations. The
   0.3\,dex spread of the various local luminosity-metallicity relations in their Fig.11
   highlights the discrepancies that can arise from different sample
   selections (theirs are fainter than ours) and metallicity calibrations
   (they use the lower branch $R_{23}$ calibration).  
They also use the log([NII]/[OII])=$-$1 as cut for the upper and lower branches and
   that the majority of their data was from McGaugh (1994).
 
  Roennback \& Bergvall (1995) 
   obtained spectroscopic observations of 16 blue LSBGs. These are among the
   bluest and dimmest objects found in the ESO-Uppsala Catalogue. 
   Oxygen abundances are derived in
   24 H~{\sc ii} regions. They are at the lower boundary found for galaxies
   with 12+log(O/H)$\sim$ 7.54-8.09. 
   These bluest, dimmest and metal-poor objects are not the same as our
   normal disk galaxies from SDSS. 

   As the discussions above, we now can understand the discrepancies 
   between the metallicity estimates
   of our LSBGs and those of the H~{\sc ii} regions in LSBGs in literature.
   It could be due to the different types of galaxies intrinsicly,
   and the different oxygen abundance calibration methods and 
   different photoionization models used.
   
\begin{figure} 
\centering
\includegraphics [width=4.1cm, height=4.0cm]{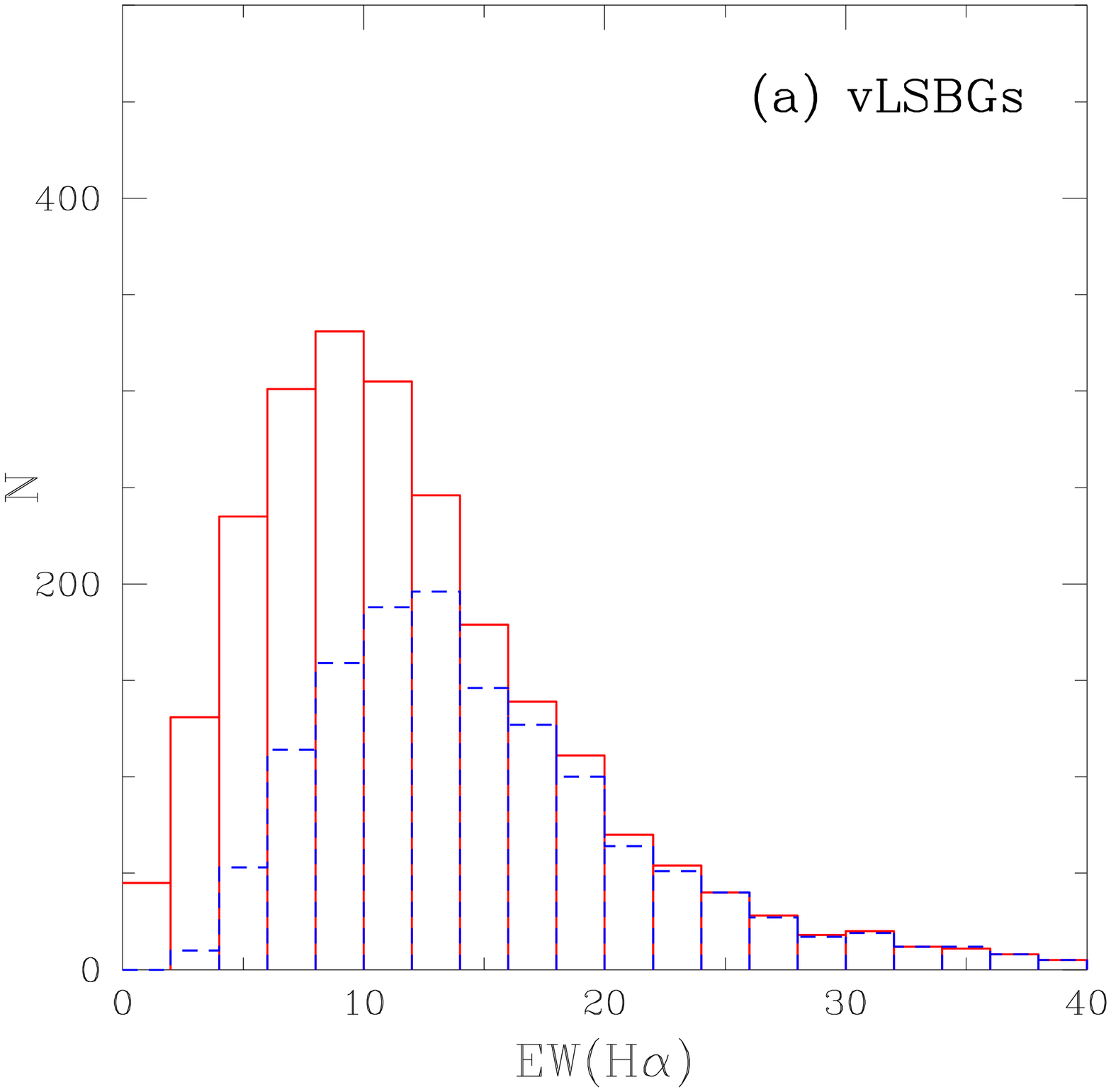} 
\includegraphics [width=4.1cm, height=4.0cm]{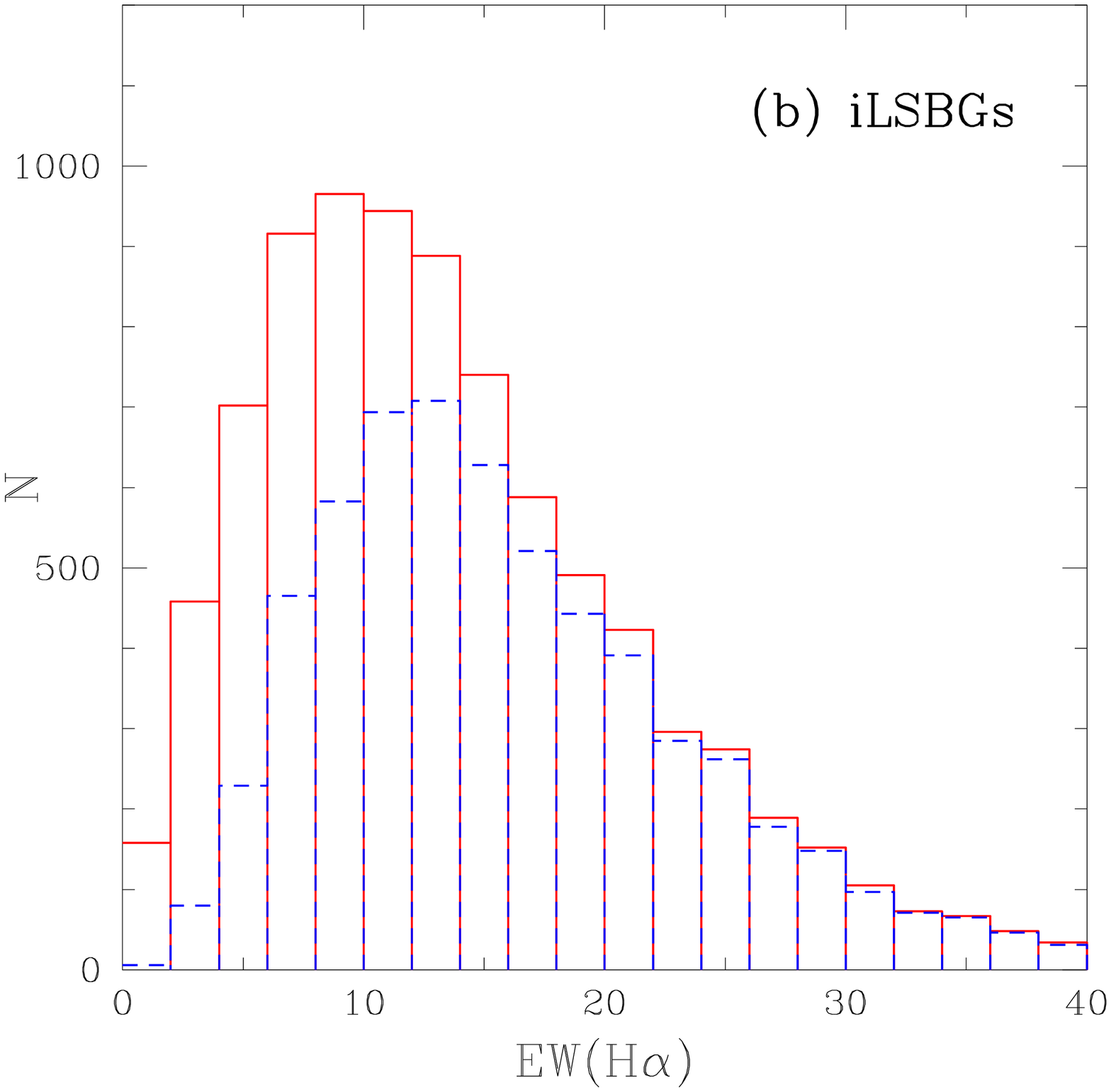}
\caption {Histogram distributions of EW(H$\alpha$) 
to compare the selected emission-line LSBGs following
criteria (i-v) in Sect.2
with those sample before considering emission-line cut 
(with criteria (i-iv) in Sect.2).
(a) for vLSBGs and (b) for iLSBGs. The solid
lines show the distribution of the sample selected before considering 
emission-line cut (2,305 and 8,637), and the dashed lines refer to those
have been considered emission-line cut (1,364 and 6,055 for vLSBGs
and iLSBGs, respectively).  
}
\label{fig.com.EWHa}
\end{figure}

    Another point we would like to address is that
   our sample galaxies selected by emission lines (see Sect.2)
   may exclude the ones without much star formation, 
   which often be as LSBGs. 
   Indeed, in all our 12,282 LSBGs (Zhong et al. 2008), there
   are 2,817 classified as vLSBGs, and the rest 9,465 are iLSBGs.
   Further with the 0.04$<z<$0.25 cut, there are 2,355 vLSBGs and 8,731 iLSBGs
   selected. Among them, 2,305 and 8,637 are measured their 
   equivalent weights of H$\alpha$ emission lines, EW(H$\alpha$), respectively. 
   Then the selected 1,364 emission-line vLSBGs (see Sect.2) correspond to about 
   59\% of those selected from photometry plus redshift cut plus
   EW(H$\alpha$). For iLSBGs, this fraction is about 70\% (=6,055/8,637).
    Fig.~\ref{fig.com.EWHa} shows the histogram distributions of 
    EW(H$\alpha$) for vLSBGs (left panel) and iLSBGs (right panel).
    Namely, the solid
lines are for the 2,305 and 8,637 sample galaxies before applying emission-line 
criterion,  
and the dashed lines are for the 1,364 and 6,055 galaxies 
of vLSBGs and iLSBGs with eimssion-line
criterion (criterion (v) in Sect.2).
They do show that most of the rejected LSBGs by emission-line cut have
lower EW(H$\alpha$) values. 

   Thus we should keep in mind
   that those galaxies without obvious emission lines 
   could have very few star formation
   and then few metal enrichment, and could have lower metal abundances then.
  We should also notice that 
  a very important point is that the previous studies target H~{\sc ii}
   regions at any position in the galaxy whereas SDSS is just sensitive
   to star formation in the nuclear region.  Thus, the previous works are
   probably include galaxies with very low SFR and just a few H~{\sc ii}
   regions that get rejected from the SDSS sample.
    
\section{Summary and conclusions}

In this second paper of our series work about a large sample of LSBGs
($\mu_0(B)>$22 mag arcsec$^{-2}$)
selected from the SDSS-DR4 main galaxy sample (Zhong et al. 2008, the Paper I),
which are low-inclination, disk-dominated galaxies, 
we study the spectroscopic properties of the star-forming 
sample galaxies including dust extinction,
strong emission-line ratios, oxygen abundances, log(N/O) abundance ratios, and 
the relations of 12+log(O/H) vs. stellar mass,  12+log(O/H) vs. $\mu_0(B)$,
and stellar mass vs. $\mu_0(B)$. 
For comparison, a large sample of HSBGs 
($\mu_0(B)<$22 mag arcsec$^{-2}$) is also selected simultaneously
and done similar analyses.
To be in more details,
the entire sample galaxies of LSBGs and HSBGs are further divided into
four subgroups according to their  $\mu_0(B)$ (in units of mag arcsec$^{-2}$):
i.e., 
vLSBGs with 24.5-22.75,
iLSBGs with 22.75-22.0,
iHSBGs with 22.0-21.25, and
vHSBGs with $<$21.25. Their resulted properties are summarized as follows.

\begin{enumerate}

\item The AGN fractions of the sample galaxies are small, less than 9\%, 
verified by the BPT diagrams from emission-line ratios.
The reason could be that our $fracDev_r$  cut has selected against 
galaxies with bulges.
We select the star-forming galaxies for further studies. 

\item 
 LSBGs span a wide range in dust
   attenuation, metallicity, N/O and stellar mass.  
   The median values of these property parameters all
   increase with surface brightness as can be seen in Table\,1.
   However, these trends can, for the most part be accounted for by the
   differences in stellar mass among the samples. 
Most of our sample galaxies have log([N~{\sc ii}]/ [O~{\sc ii}])$>-$1.2,
which means they belong to the upper branch of 
$R_{23}$ vs. metallicity relation. 
Thus our vLSBGs and iLSBGs are not as metal-poor
as the H~{\sc ii} regions in a small sample of LSBGs studied 
by McGaugh (1994) and some following works
(8.06-8.20 of 12+log(O/H) in median).
One reason of this discrepancy could be that our sample galaxies are 
more luminous than theirs intrinsicly.
The second reason could be the different calibrations of $R_{23}$ 
for metallicity, previous studies, such as McGaugh et al. (1994), 
adopted log([N~{\sc ii}]/ [O~{\sc ii}])$>-$1.0 (higher than $-1.2$, what we used) 
to judge and found most of their samples were in the lower branch of 
$R_{23}$ vs. metallicity relation, 
while we adopt the oxygen abundances of the galaxies obtained by
using the Bayesian approach by the MPA/JHU group as Tremonti et al. (2004).
This could be well explained by Fig.~\ref{fig.com.oh.lower} for 
those our sample
galaxies having $-$1.2$<$log([N~{\sc ii}]/[O~{\sc ii}])$<-$1.0. 

\item
The log(N/O) abundances
of our sample galaxies
are more consistent with the combination of the ``primary" and ``secondary"
components, but the ``secondary" component dominates. 
In the log(N/O) vs. 12+log(O/H) relations of the four subgroup galaxies,
the medain values of them at given O/H 
show slight differences ($<$0.074\,dex) following surface brightness
as shown in Fig.~\ref{fig.no.oh} and Fig.~\ref{fig.dis.on.referee}a. 
These may mean that
the sample galaxies may have some but not quite much 
difference in star formation history,
stellar populations,
and enrichment history of chemical elements, 
such as nitrogen and oxygen, as well as the ratio
of intermediate-mass to massive stars.

\item
The relations between stellar masses and 12+ log(O/H) for our sample galaxies show that
the vLSBGs have less metal-rich and massive ones than other three subgroups, 
And the iLSBGs have also less such objects than the iHSBGs and vHSBGs.
 We would like to highlight that 
  the LSBGs in our sample span a wide range of stellar mass and metallicity
  although having less massive and metal-rich ones;
  LSBGs are not much more metal poor than HSBGs of the same mass
 (less than 0.11\,dex shown by the median-value points in Fig.~\ref{fig.dis.on.referee}b).
 
\item 
 For the vLSBGs and iLSBGs, the galaxies with lower surface
 brightnesses have lower metallicities,
 and have smaller stellar masses generally.
 The iHSBGs also follow this trend
 slightly, but the vHSBGs do not. 
 These trends could be intrinsicly linked to
 stellar masses. 
 If the effects of stellar masses are
 removed by calculating the 
 residuals between the measured oxygen abundances and
 those calculated from stellar masses using formula,
 then the four subgroups do not show much difference
 (from +0.035 to -0.039\,dex as given in Fig.~\ref{fig.mu.oh.mass}c).
 In a word, the
 metallicity of LSBGs is correlated with $\mu_0(B)$, but more
 tightly correlated with stellar mass; 
 the apparent correlation of $\mu_0(B)$ and O/H is a consequence of the
 correlation with stellar mass; 
 the residuals of the MZR show very slightly
 correlation with $\mu_0(B)$, 
 which means that
 these LSBGs and HSBGs may have some variation but not quite much 
 in star formation history
 and chemical enrichment history.  
 
\end{enumerate}
 
\section*{Acknowledgments}
We appreciate our referee, who
provide very admirable, constructive and helpful comments
and suggestions, which help to improve well our work.
We would like to thank Prof. Rob Kennicutt and Myriam Rodrigues 
for helpful and interesting discussions 
during the IAU Symposium 262 in Rio de Janeiro in August 2009. 
We thank Dr. Shiyin Shen and Dr. Ruixiang Chang for useful discussions.
 This work was supported by the Natural Science Foundation of China
 (NSFC) Foundation under Nos.10933001, 10973006,
10973015, 10673002; and the
 National Basic Research Program of China (973 Program) Nos.2007CB815404,
 2007CB815406, and No. 2006AA01A120 (863 project).
 We thank the SDSS and the MPA/JHU group to make the database and
 many measurements of the SDSS galaxies be available for public use.

\label{lastpage}

\end{document}